\documentclass[12pt]{article}
\usepackage{makeidx}
\usepackage{amssymb}
\usepackage{amsfonts}
\usepackage{amsmath}
\usepackage{graphicx}
\usepackage{setspace}
\usepackage{authblk}
\usepackage{units}
\usepackage{appendix}
\usepackage[left=2cm,top=2cm,right=2cm]{geometry}

\begin{document}

\title{\textbf{Coulomb field in a constant electromagnetic background}}
\author[1]{T. C. Adorno\thanks{tg.adorno@mail.tsu.ru, tg.adorno@gmail.com}}
\author[1,2,3]{D. M. Gitman\thanks{gitman@if.usp.br}}
\author[1,2]{A. E. Shabad\thanks{shabad@lpi.ru}}
\affil[1]{\textit{Department of Physics, Tomsk State University, Lenin Prospekt 36, 634050, Tomsk, Russia;}}
\affil[2]{\textit{P. N. Lebedev Physical Institute, 53 Leninskiy prospekt,
119991, Moscow, Russia;}}
\affil[3]{\textit{Instituto de F\'{\i}sica, Universidade de S\~{a}o Paulo, Caixa Postal 66318, CEP 05508-090, S\~{a}o Paulo, S.P., Brazil;}}

\maketitle

\onehalfspacing

\begin{abstract}
Nonlinear Maxwell equations are written up to the third-power deviations
from a constant-field background, valid within any local nonlinear
electrodynamics including QED with a Euler-Heisenberg (EH) effective
Lagrangian. The linear electric response to an imposed static finite-sized charge
is found in the vacuum filled by an arbitrary combination of constant and
homogeneous electric and magnetic fields. The modified Coulomb field and
corrections to the total charge and to the charge density are given in terms
of derivatives of the effective Lagrangian with respect to the field
invariants. These are specialized for the EH Lagrangian. 
\end{abstract}

\newpage

\section{Introduction\label{Sec1}}

The large values of electromagnetic fields, close to or larger than $\frac{%
m^{2}}{e}=4.4\cdot 10^{13}G=1.6\cdot 10^{16}V/cm$, are present not only in
astrophysical objects, such as pulsars, magnetars and quark stars, but also
in the close vicinities of elementary particles as produced by their
charges, magnetic and electric dipole moments. (For instance, the magnetic
field of the neutron at the edge of its electromagnetic radius makes up the
value of about $10^{16}G$, characteristic of magnetars.) This fact is
encouraging interest in nonlinear electrodynamics that accounts for the
effects of interaction between electromagnetic fields when at least one of
them is large. These are, for instance, linear, quadratic and cubic
responses of the vacuum, that include large \textquotedblleft
background\textquotedblright\ or \textquotedblleft
external\textquotedblright\ electromagnetic fields, to probe fields of, say,
the ones produced by charges and currents. Another example of nonlinear
effects is the correction to the magnetic and electric dipole moments of
elementary particles and resonances owing to their self-coupling.

Another circumstance that has been fueling interest in the effects of
quantum electrodynamics, this time within the framework of (2+1)-dimensional
space-time, is the problem of Coulomb impurities in graphene; see the most
recent review in Ref \cite{Kotov} and also the recent paper \cite{GavGitYok202012}.
The point is that the effective coupling in that theory is much larger, and,
correspondingly, the values of the fields optimal for nonlinearity are much
smaller than in the (3+1)-dimensional QED.

Among the possible background fields most handy are those that admit an exact
solution to the Dirac equation for the electron, which enables explicitly
exploiting the Furry picture in order to take the background exactly while
calculating the response functions -- the polarization operators of different
ranks. These are the plane-wave field (see the review \cite{laser}), the
Coulomb field (dealt with in Ref. \cite{MilStra1983}), and the constant
homogeneous field \cite{Shabad book, Dittrich}. However, only the latter
admits, strictly speaking, a treatment with the use of a local electromagnetic
action. The local approach, thanks to its relative simplicity, proved to
be very fruitful in revealing special nonlinear effects \cite{ShaGit2012}.
These include the consequences of nonlinearity of quantum electrodynamics
stemming from the quantum phenomenon of virtual electron-positron pairs
creation by a photon, the pairs interacting with an electromagnetic field
before their mutual annihilation. The corresponding interaction between
electromagnetic fields is included at the local level by appealing to the
Euler-Heisenberg effective action either at the one-loop \cite{EH, Weiskopf,
BerLifPit} or two-loop \cite{Ritus2loop} level giving rise to a modification
of the classical Maxwell equations. Among the other, intrinsically nonlinear
local classical theories of electromagnetism, the most popular is the
geometrically elegant Born-Infeld model \cite{BorInf34} , also arisen in the
low-energy limit out of the string theory in the electromagnetic sector \cite%
{FraTse}. Its extensions \cite{GaHe, Kruglov}\ are also in use. All these
models are automatically included in our local treatment.

Many effects of nonlinear electrodynamics were considered before. The ones
for which the three-photon diagram in an external magnetic field is responsible
are the photon splitting \cite{Adler} and the quadratic in the electric
charge magnetoelectric effect when the charge becomes a magnetic dipole \cite%
{AdoGitSha2014}. The cubic effect for which the light-by-light four-photon
diagram is responsible was considered \cite{CaiGitSha2013} without a
background to produce nonlinear corrections to electric and magnetic dipoles
and other sources. In the present paper, we are studying solutions of the
Maxwell equations linearized above the background, which is an arbitrary
combination of space- and time-independent electric and magnetic fields. We
concentrate on finding the electric response to the applied field of a static
charge; in other words, we are looking for the Coulomb field modified by the
presence of this complex background. Previously, the Coulomb correction was
considered (also beyond the local regime) in the external magnetic field
taken alone \cite{shabus,SadSod07,13th Lomonosov,MacVys11,GodVys}. In the
combined field considered here, also a magnetic response to the electrostatic
source exists. The preliminary study of it has indicated the possible
presence of a solution carrying a magnetic monopole \cite{magncharge} in the
simplifying case of a small background and parallel background fields.
However, we postpone the detailed description of the magnetic response in
the general case free of these restrictions to a forthcoming paper.

In Sec. \ref{Sec2}, supported by Appendix \ref{Ap1}, we present the
derivation of nonlinear Maxwell equations with background fields including
quadratic and cubic nonlinearities. In Sec. \ref{Sec2}, we find the linear
correction to the Coulomb field in a constant background. In Subsec. \ref%
{charge}, we describe the current induced by the background fields and by the
probe field of an electric charge of finite radius, and we find the
nonvanishing correction to the total charge. In Subsec. \ref{field}, we study the solutions for the modified Coulomb field calculated in Appendix %
\ref{AppII} by using the projector method that separates the solutions
subject to the first pair of Maxwell equations. The special cases of parallel
external fields and the corresponding scalar potential are considered in
Secs. \ref{parallel} and \ref{potential}.\emph{\ }In Sec. %
\ref{EH}, we concentrate on some realizations of the considered effects
stemming from the use of the Euler-Heisenberg Lagrangian as a quantum source
of nonlinearity. We consider small- and large-field limits, especially an
extension of the effect of screening of the charge by strong magnetic field \cite%
{shabus,SadSod07,13th Lomonosov,MacVys11,GodVys} to the case where a smaller electric field is also present. The results are finally discussed in the conclusions.

\section{Nonlinear Maxwell equations at the infrared regime\label{Sec2}}

\subsection{Maxwell equations for arbitrary strong fields}

The equations of motion of a nonlinear Maxwell theory, which incorporates
the contribution of an effective field theory into its formulation, are
extracted from the action\footnote{%
Greek indices span the 4-dimensional Minkowski space-time taking the values
0,1,2,3. The Minkowski metric tensor has the signature $\eta _{\mu \nu }=%
\mathrm{diag}(+1,-1,-1,-1)$ and bold symbols are reserved for
three-dimensional Euclidean vectors (for instance $\mathbf{A}(x)=\left(
A^{i}(x)\right) \,,\,\,i=1,2,3)$. The Heaviside system of units where $\hbar
=c=1$ (the fine structure constant being $\alpha =e^{2}/4\pi $) is used
throughout the paper. The four-rank antisymmetric tensor has the
normalization $\varepsilon ^{0123}=+1.$},%
\begin{eqnarray}
&&S\left[ A\right] =S_{\mathrm{Max}}\left[ A\right] +S_{\mathrm{int}}\left[ A%
\right] +\Gamma \left[ A\right] \,,  \notag \\
&&S_{\mathrm{Max}}\left[ A\right] =-\int \mathfrak{F}\left( x\right)
d^{4}x\,,\ \ S_{\mathrm{int}}\left[ A\right] =-\int J_{\mu }\left( x\right)
A^{\mu }\left( x\right) d^{4}x\,,\ \ \Gamma \left[ A\right] =\int \mathfrak{L%
}\left( x\right) d^{4}x\,,  \label{action}
\end{eqnarray}%
in the form of Euler-Lagrange equations $\delta S\left[ A\right] /\delta
A^{\mu }\left( x\right) $,%
\begin{equation}
\partial ^{\nu }F_{\nu \mu }\left( x\right) +\frac{\delta \Gamma \left[ A%
\right] }{\delta A^{\mu }\left( x\right) }=J_{\mu }\left( x\right) \,,
\label{nm3}
\end{equation}%
where $A^{\mu }\left( x\right) $ are electromagnetic potentials, $S_{\mathrm{%
Max}}\left[ A\right] $ is the free Maxwell action,\ $S_{\mathrm{int}}\left[ A%
\right] $ is the interaction with a classical current $J_{\mu }\left(
x\right) $, and $\Gamma \left[ A\right] $ and $\mathfrak{L}\left( x\right) $ are
the effective action and Lagrangian, respectively. Here $F^{\mu \nu }\left(
x\right) =\partial ^{\mu }A^{\nu }(x)-\partial ^{\nu }A^{\mu }(x)$ is the
usual electromagnetic field strength, $\tilde{F}^{\mu \nu }\left( x\right)
=\left( 1/2\right) \varepsilon ^{\mu \nu \rho \sigma }F_{\rho \sigma }\left(
x\right) $ is its dual and $\mathfrak{F}$, $\mathfrak{G}$ stand for the
field invariants $\mathfrak{F}\left( x\right) =\frac{1}{4}F^{\mu \nu }\left(
x\right) F_{\mu \nu }\left( x\right) $ and $\mathfrak{G}\left( x\right)
=\left( 1/4\right) \tilde{F}^{\mu \nu }\left( x\right) F_{\mu \nu }\left(
x\right) $, respectively. Through the identities,%
\begin{eqnarray}
\frac{\delta \Gamma \left[ A\right] }{\delta A^{\mu }\left( x\right) }
&=&\int dz\left( \frac{\delta \mathfrak{L}\left( \mathfrak{F},\mathfrak{G}%
\right) }{\delta \mathfrak{F}\left( z\right) }F_{\tau \mu }\left( z\right) +%
\frac{\delta \mathfrak{L}\left( \mathfrak{F},\mathfrak{G}\right) }{\delta 
\mathfrak{G}\left( z\right) }\tilde{F}_{\tau \mu }\left( z\right) \right) 
\frac{\partial }{\partial z_{\tau }}\delta ^{4}\left( x-z\right) \,,  \notag
\\
\frac{\delta \mathfrak{F}\left( z\right) }{\delta F^{\alpha \beta }\left(
x\right) } &=&\frac{1}{2}F_{\alpha \beta }\left( z\right) \delta \left(
x-z\right) \,,\ \ \frac{\delta \mathfrak{G}\left( z\right) }{\delta
F^{\alpha \beta }\left( x\right) }=\frac{1}{2}\tilde{F}_{\alpha \beta
}\left( z\right) \delta \left( x-z\right) \,,  \label{nm3.1}
\end{eqnarray}%
the Euler-Lagrange equations (\ref{nm3}) become the nonlinear Maxwell
equations at the infrared limit,%
\begin{equation}
\partial ^{\nu }F_{\nu \mu }\left( x\right) -\partial ^{\tau }\left[ \frac{%
\delta \mathfrak{L}\left( \mathfrak{F},\mathfrak{G}\right) }{\delta 
\mathfrak{F}\left( x\right) }F_{\tau \mu }\left( x\right) +\frac{\delta 
\mathfrak{L}\left( \mathfrak{F},\mathfrak{G}\right) }{\delta \mathfrak{G}%
\left( x\right) }\tilde{F}_{\tau \mu }\left( x\right) \right] =J_{\mu
}\left( x\right) \,,  \label{MaxEq}
\end{equation}%
assuming that the action (\ref{action}) is a Lorentz-invariant functional
only of the field strengths, and not on potentials $A^{\mu }\left( x\right) $
due to the gauge invariance. These equations are nonlinear, because the
unknown fields enter into the functional derivatives above in a complicated
way. The second term in this equation may be referred to as a nonlinearly
induced current:%
\begin{equation}
j_{\mu }^{\text{\textrm{nl}}}=\partial ^{\tau }\left( \frac{\delta \mathfrak{%
L}\left( \mathfrak{F},\mathfrak{G}\right) }{\delta \mathfrak{F}\left(
x\right) }F_{\tau \mu }\left( x\right) +\frac{\delta \mathfrak{L}\left( 
\mathfrak{F},\mathfrak{G}\right) }{\delta \mathfrak{G}\left( x\right) }%
\tilde{F}_{\tau \mu }\left( x\right) \right) \,.  \label{nonlincurrent}
\end{equation}%
The nonlinear Maxwell equations above must be completed with the first pair
of Maxwell equations%
\begin{equation}
\partial _{\nu }\widetilde{F}^{\nu \mu }\left( x\right)=0
\label{bianchi}
\end{equation}%
(which, indeed, is an identity of the classical electrodynamics formulated
in terms of potentials, instead of field strengths.) which are to be
additionally postulated, in the same way as they were in the linear
Faraday-Maxwell theory on the basis of the apparent absence of a magnetic
charge from nature. Although we have pointed \ a nontrivial magnetic charge
under certain conditions \cite{magncharge}, internally induced within the
nonlinear theory, in the present paper we are dealing only with solutions
obeying Eq. (\ref{bianchi}), while leaving other solutions of Eqs. (\ref%
{MaxEq}) for the upcoming paper.

\emph{Remark}. Two exact solutions to Eqs. (\ref{MaxEq}) and (\ref{bianchi}%
) with zero source $J_{\mu }\left( x\right) =0$\ are known. One is quite
evident. It is the field $F_{\alpha \beta }$ independent of the space-time
coordinate $x_{\mu }$. Another exact solution with the zero source is the
plane-wave laser field. This statement is equivalent to the one derived in Ref. 
\cite{FradSha} appealing also to the $U(1)$ gauge and charge invariance. It
is associated with the infrared problem in a massless vector theory. There
is also a third, very exotic, nilpotent solution to the pair of exact field
equations (\ref{MaxEq}) and (\ref{bianchi}) in quantum electrodynamics owing to
the absence of the asymptotic freedom in it that leads to spontaneous
breakdown of translational invariance \cite{FradSha}.

From (\ref{MaxEq}), below in this section we derive nonlinear Maxwell
equations for small deviations from a certain background field. When doing
it we keep to the infrared limit, that describes slowly varying
electromagnetic fields. In this case one takes the effective action $\Gamma %
\left[ A\right] $ as a local functional of the field invariants $\mathfrak{F}
$ and $\mathfrak{G}$. This means that the effective Lagrangian $\mathfrak{L}=%
\mathfrak{L}\left( \mathfrak{F},\mathfrak{G}\right) $ in (\ref{action}) does
not include space-time derivatives of the field intensities.
Correspondingly, partial derivatives should be substituted in (\ref{MaxEq})
in the place of functional ones.

\subsection{Expansion of the nonlinear Maxwell equations around a background
field\label{Expansion}}

In what follows we divide the current into two parts $J_{\mu }\left(
x\right) =\mathcal{J}_{\mu }\left( x\right) +j_{\mu }\left( x\right) $ and
treat the field strength as a sum $F^{\alpha \beta }\left( x\right) =%
\mathcal{F}^{\alpha \beta }\left( x\right) +f^{\alpha \beta }\left( x\right) 
$ of a background field $\mathcal{F}^{\alpha \beta }\left( x\right) $,
provided by the sources $\mathcal{J}_{\mu }\left( x\right) $, and a small
deviation $f^{\alpha \beta }\left( x\right) $ from it owing to $j_{\mu
}\left( x\right) $. In other words, the background field is subjected to the
nonlinear equation%
\begin{equation}
\partial ^{\nu }\mathcal{F}_{\nu \mu }\left( x\right) -\partial ^{\nu }\left[
\left. \frac{\delta \mathfrak{L}\left( \mathfrak{F},\mathfrak{G}\right) }{%
\delta \mathfrak{F}\left( x\right) }\right\vert _{F=\mathcal{F}}\mathcal{F}%
_{\nu \mu }\left( x\right) +\left. \frac{\delta \mathfrak{L}\left( \mathfrak{%
F},\mathfrak{G}\right) }{\delta \mathfrak{G}\left( x\right) }\right\vert _{F=%
\mathcal{F}}\widetilde{\mathcal{F}}_{\nu \mu }\left( x\right) \right] =%
\mathcal{J}_{\mu }\left( x\right) \,,  \label{background}
\end{equation}%
where it is indicated that the field invariants $\mathfrak{F,G}$ are formed
by the background field.

To find equations for the deviation $f^{\alpha \beta }\left( x\right) $ it
is needed to expand Eq. (\ref{MaxEq}) in its powers, with the use of
Eq. (\ref{background}) for the zero-order term. By truncating these
expansions at any power of $f^{\alpha \beta }\left( x\right) $, equations
with the polynomial nonlinearity are constructed. In Appendix \ref{Ap1} we
present the tensor-valued coefficients in the expansion of the quantities $%
\frac{\delta \mathfrak{L}\left( \mathfrak{F},\mathfrak{G}\right) }{\delta 
\mathfrak{F}\left( x\right) }F_{\tau \mu }\left( x\right) $ and$\ \frac{%
\delta \mathfrak{L}\left( \mathfrak{F},\mathfrak{G}\right) }{\delta 
\mathfrak{G}\left( x\right) }\tilde{F}_{\tau \mu }\left( x\right) $,
involved in Eq. (\ref{MaxEq})\ in powers of the deviations $f^{\alpha
\beta }\left( x\right) $ from the background field up to the third power%
\footnote{%
Although in the present paper we deal only with a theory with parity
conservation, the formulas of Appendix \ref{Ap1} are written so as to
include the possible violation of parity.}. Their substitution into (\ref%
{MaxEq}) is rather cumbersome, although it is simplified if certain
restrictions are imposed on the background field $\mathcal{F}_{\mu \nu
}\left( x\right) $. However as far as the first power of the deviation is
concerned, all terms shall be kept in subsequent calculations.

We label derivatives of the effective Lagrangian as follows%
\begin{eqnarray}
&&\left. \frac{\partial \mathfrak{L}}{\partial \mathfrak{F}}\right\vert _{F=%
\mathcal{F}}=\mathfrak{L}_{\mathfrak{F}}\,,\ \ \left. \frac{\partial 
\mathfrak{L}}{\partial \mathfrak{G}}\right\vert _{F=\mathcal{F}}=\mathfrak{L}%
_{\mathfrak{G}}\,,  \notag \\
&&\left. \frac{\partial ^{2}\mathfrak{L}}{\partial \mathfrak{F}^{2}}%
\right\vert _{F=\mathcal{F}}=\mathfrak{L}_{\mathfrak{FF}}\,,\ \ \left. \frac{%
\partial ^{2}\mathfrak{L}}{\partial \mathfrak{F}\partial \mathfrak{G}}%
\right\vert _{F=\mathcal{F}}=\mathfrak{L}_{\mathfrak{FG}}\,,\ \ \left. \frac{%
\partial ^{2}\mathfrak{L}}{\partial \mathfrak{G}^{2}}\right\vert _{F=%
\mathcal{F}}=\mathfrak{L}_{\mathfrak{GG}}\,.  \label{nm8.0}
\end{eqnarray}%
Then, using Eqs. (\ref{nm5.3}), (\ref{nm7.1}) from Appendix \ref{Ap1}, we
get the linearized equation (\ref{MaxEq}),%
\begin{eqnarray}
\partial ^{\nu }f_{\nu \mu }\left( x\right) &=&j_{\mu }^{\mathrm{lin}}\left(
x\right) +j_{\mu }\left( x\right) \,,  \notag \\
j_{\mu }^{\mathrm{lin}}\left( x\right) &=&\partial ^{\tau }\left[ \mathfrak{L%
}_{\mathfrak{F}}f_{\tau \mu }\left( x\right) +\frac{1}{2}\left( \mathfrak{L}%
_{\mathfrak{FF}}\mathcal{F}_{\alpha \beta }+\mathfrak{L}_{\mathfrak{FG}}%
\widetilde{\mathcal{F}}_{\alpha \beta }\right) \mathcal{F}_{\tau \mu
}f^{\alpha \beta }\left( x\right) \right]  \notag \\
&+&\partial ^{\tau }\left[ \mathfrak{L}_{\mathfrak{G}}\tilde{f}_{\tau \mu
}\left( x\right) +\frac{1}{2}\left( \mathfrak{L}_{\mathfrak{FG}}\mathcal{F}%
_{\alpha \beta }+\mathfrak{L}_{\mathfrak{GG}}\widetilde{\mathcal{F}}_{\alpha
\beta }\right) \widetilde{\mathcal{F}}_{\tau \mu }f^{\alpha \beta }\left(
x\right) \right] \,,  \label{nm8.1}
\end{eqnarray}%
for the linear response $f_{\nu \mu }\left( x\right) $ to the small current $%
j_{\mu }\left( x\right) $ in the presence of background field $\mathcal{F}%
_{\alpha \beta }$, produced by the current $\mathcal{J}_{\mu }\left(
x\right) $ via Eq. (\ref{background}). Here and below the subscripts by $%
\mathfrak{L}$ denote the multiple derivatives of the effective Lagrangian
with respect the field invariants $\mathfrak{F}$ and $\mathfrak{G}$\ with
the background field substituted into them after the derivatives are
calculated. When the background is the constant field of the most general
form, we have $\mathcal{J}_{\mu }\left( x\right) =0$ and the coefficient
functions $\mathfrak{L}_{\mathfrak{F}\mathcal{\ }},$ $\mathfrak{L}_{%
\mathfrak{FF}\mathcal{\ }},$ $\mathfrak{L}_{\mathfrak{FG}},$ $\mathfrak{L}_{%
\mathfrak{G}},$ $\mathfrak{L}_{\mathfrak{GG}}$ are not subject to the
space-time differentiation indicated in (\ref{nm8.1}).

For the special case where the background field has its second invariant $%
\mathfrak{G}$ equal to zero, setting $\mathfrak{G=}$ $0$ implies $\mathfrak{L%
}_{\mathfrak{FG}}=$ $\mathfrak{L}_{\mathfrak{FFG}}=0$ due to parity
conservation, since the latter requires that the Lagrangian $\mathfrak{L}$
be even in $\mathfrak{G}$. In this case the quadratic contribution from (\ref%
{nm7.2}), (\ref{nm5.4}), (\ref{nm7.1}) to the nonlinearly induced current (%
\ref{nonlincurrent}) can be split in two parts,%
\begin{equation*}
j_{\mu }^{\text{\textrm{nl}}}(x)=j_{\mu }^{\mathrm{lin}}\left( x\right)
+j_{\mu }^{\text{\textrm{qua}}}\left( x\right) \,,
\end{equation*}%
where $j_{\mu }^{\mathrm{lin}}\left( x\right) $ has been written as Eq. (\ref%
{nm8.1}) and $j_{\mu }^{\text{\textrm{qua}}}\left( x\right) $ represents the
quadratic in the deviation part, expressed as%
\begin{equation}
j_{\mu }^{\text{\textrm{qua}}}(x)=\frac{1}{8}\frac{\partial }{\partial
x_{\alpha }}\mathcal{O}_{\mu \tau \sigma \alpha \beta \gamma }f^{\beta \tau
}f^{\gamma \sigma },  \label{nonlincur2}
\end{equation}%
where%
\begin{align}
& \mathcal{O}_{\mu \tau \sigma \alpha \beta \gamma }=\mathfrak{L_{GG}}\left[ 
\widetilde{\mathcal{F}}_{\gamma \sigma }\epsilon _{\alpha \mu \beta \tau }+%
\widetilde{\mathcal{F}}_{\alpha \mu }\epsilon _{\beta \tau \gamma \sigma }+%
\widetilde{\mathcal{F}}_{\beta \tau }\epsilon _{\alpha \mu \gamma \sigma }%
\right] +  \notag \\
& +\mathfrak{L_{FF}}\left[ \left( \eta _{\mu \tau }\eta _{\alpha \beta
}-\eta _{\mu \beta }\eta _{\alpha \tau }\right) \mathcal{F}_{\gamma \sigma }+%
\mathcal{F}_{\alpha \mu }\left( \eta _{\tau \sigma }\eta _{\gamma \beta
}-\eta _{\beta \sigma }\eta _{\gamma \tau }\right) +\mathcal{F}_{\beta \tau
}\left( \eta _{\mu \sigma }\eta _{\gamma \alpha }-\eta _{\alpha \sigma }\eta
_{\gamma \mu }\right) \right]  \notag \\
& +\mathfrak{L_{FGG}}\left[ {\mathcal{F}}_{\alpha \mu }\widetilde{\mathcal{F}%
}_{\beta \tau }\widetilde{\mathcal{F}}_{\gamma \sigma }+\widetilde{\mathcal{F%
}}_{\alpha \mu }\mathcal{F}_{\beta \tau }\widetilde{\mathcal{F}}_{\gamma
\sigma }+\widetilde{\mathcal{F}}_{\alpha \mu }\widetilde{\mathcal{F}}_{\beta
\tau }\mathcal{F}_{\gamma \sigma }\right] +\mathfrak{L_{FFF}}\mathcal{F}%
_{\alpha \mu }\mathcal{F}_{\beta \tau }\mathcal{F_{\gamma \sigma }}\,.
\label{sixrank}
\end{align}%
The current (\ref{nonlincur2}) may be thought of as corresponding to the
three-photon diagram (photon splitting/merging) beyond the photon mass
shell, nonzero against a nontrivial background \cite{Birula}. In our past
papers \cite{ShaGit2012,AdoGitSha2014} we studied the corresponding
nonlinear equation quadratic in the field resulted from the truncation at
the second power, but with $\mathfrak{G}=0$. The quadratic response to an
applied Coulomb field of an electric charge was considered when the
background field is a constant magnetic field $\mathfrak{F}>0,$ $\mathfrak{G}%
=0$ in the rest frame of the charge. Once the background field is constant\
the matrix (\ref{sixrank}) can be taken from under the derivative sign in (%
\ref{nonlincur2}). It was found that the response was purely magnetic and
formed a magnetic dipole.

In Ref. \cite{CaiGitSha2013} (see also the Eq. (\ref{nm9.4}) in Appendix \ref%
{Ap1}) we considered the third-power nonlinearity, but with no background.
The cubic contribution to the nonlinearly induced current (\ref%
{nonlincurrent}) is%
\begin{equation}
j_{\mu }^{\text{\textrm{nl}}}\simeq \frac{1}{4}\mathfrak{L}_{\mathfrak{FF}}%
\frac{\partial }{\partial x_{\tau }}f_{\tau \mu }\left( x\right) f_{\alpha
\beta }\left( x\right) f^{\alpha \beta }\left( x\right) +\frac{1}{4}%
\mathfrak{L}_{\mathfrak{GG}}\frac{\partial }{\partial x_{\tau }}\tilde{f}%
_{\tau \mu }\left( x\right) f^{\alpha \beta }\left( x\right) \tilde{f}%
_{\alpha \beta }\left( x\right) .  \label{jnl}
\end{equation}%
This is deduced from (\ref{nm5.5}) and (\ref{nm7.2}) by setting $\mathcal{F}%
=0$ and $\mathfrak{L}_{\mathfrak{FG}}=0$. Here it is meant that the
background field invariants $\mathfrak{F}$ and $\mathfrak{G}$ are set equal
to zero after the derivatives $\mathfrak{L}_{\mathfrak{FF}}$ and $\mathfrak{L%
}_{\mathfrak{GG}}$ are calculated. It is directly seen from (\ref{sixrank})
that in the no-background case $\mathcal{F}=\widetilde{\mathcal{F}}=0$ the
quadratic current (\ref{nonlincur2}) vanishes, reflecting the vanishing of
the three-photon diagram due to the charge invariance. The remaining cubic
self-coupling of electromagnetic field introduced by the current (\ref{jnl})
(light-by-light scattering beyond the photon mass shell) leads to
corrections to static fields of electric charges and magnetic moments, as
described in Ref. \cite{CaiGitSha2013} (see also \cite{Shishmarev}). If
taken seriously for short distances from a point charge, this selfcoupling
results in finiteness of its field energy \cite{CaiGitSha2015}. The interaction
between two collinear (colliding) free electromagnetic fields was studied in
Ref. \cite{King} against the blank background using terms also of higher
power than in (\ref{jnl}).

In the present paper we restrict ourselves to effects of the first order, i.
e. we shall handle Eqs. (\ref{nm8.1}) with the constant electric and
magnetic field of arbitrary strength and mutual orientation, i. e. with $%
\mathfrak{G}\neq 0$, in contrast to the references mentioned above in this
subsection. The current $j_{\mu }\left( x\right) $ will be that of a static
electric charge of finite extension. In principle, a better treatment for
approaching the same problem, free of the limitations to slowly varying
solutions, is possible, based on the known expressions of the polarization
operator in constant external electric and magnetic fields of the most
general form, calculated beyond the present infrared (local) approximation
in Ref. \cite{BatSha1971}. However, these expressions are far less transparent
and up to now only the magnetic response was considered in some detail \cite%
{ShaUs2010}. Besides, as long as a finite-sized source is concerned the
infrared approximation is sufficient.

\section{Linear vacuum response to applied electrostatic source in
constant-field background}

\subsection{General part}

Henceforward we shall treat Eq. (\ref{nm8.1}) by perturbations. This is
meaningful as long as the coefficients in this equation are considered as
small. For instance, if the underlying nonlinear theory (\ref{action}) is
quantum electrodynamics with the effective action $\Gamma \left[ A\right] $
being the generating functional of one-electron-irreducible vertex functions
(see \cite{weinberg}), the quantities $\mathfrak{L}_{\mathfrak{F}}$, $%
\mathfrak{L}_{\mathfrak{G}}$, $\mathfrak{L}_{\mathfrak{FF}}$, and $\mathfrak{L}_{%
\mathfrak{FG}}\mathfrak{\ }$are\ proportional to the fine-structure constant 
$\alpha $ (besides being functions of the background field, wherein the
latter enters multiplied by the electron charge $e$). In what follows all
quantities associated with the constant field will be represented with a bar
above. For instance, the constant electric and magnetic components of $%
\overline{\mathcal{F}}_{\alpha \beta }$ are $\overline{E}^{i}=\overline{%
\mathcal{F}}_{0i}$, and $\overline{\mathcal{F}}_{ij}=-\varepsilon _{ijk}%
\overline{B}^{k}$. It should be stressed that after the charge distribution
is fixed the problem is deprived of the relativistic invariance.
Henceforward all the fields are understood to be attributed to the reference
frame where the charge distribution is defined.

We are going to consider perturbations of the vacuum with the constant
background field $\overline{\mathcal{F}}_{\alpha \beta }$ in response to a
small static current $j_{\mu }\left( x\right) $%
\begin{eqnarray}
&&j_{\mu }\left( x\right) =\delta _{\mu 0}\rho ^{\left( 0\right) }\left(
r\right) \,,\text{ }r=|\mathbf{x|\,,}  \notag \\
&&\rho ^{\left( 0\right) }\left( r\right) =\frac{3q}{4\pi R^{3}}\theta
\left( R-r\right) \,,\ \ R=\mathrm{const.}\,,  \label{const charge}
\end{eqnarray}%
which is a constant electric charge $q$ homogeneously distributed within a
sphere of the radius $R$, to be referred to below also as a core. In (\ref%
{const charge}), $\theta \left( z\right) $ is the Heaviside step function.
Here $q$ is the algebraic charge such that $q=-e$, for an electron, where $%
e>0$ is the absolute value of the electric charge.

In the zeroth order in $\alpha $, Eq. (\ref{nm8.1}) becomes the ordinary
Maxwell equation%
\begin{equation}
\partial ^{\nu }f_{\nu \mu }^{\left( 0\right) }\left( x\right) =j_{\mu
}\left( x\right) \,,  \label{nm8.1.1}
\end{equation}%
and we take the ordinary regularized Coulomb field%
\begin{eqnarray}
&&f_{0i}^{\left( 0\right) }\left( \mathbf{x}\right) =E^{\left( 0\right)
i}\left( \mathbf{x}\right) =\frac{q}{4\pi }\Phi \left( r\right) x^{i}\,,
\label{in2} \\
&&\Phi \left( r\right) =\frac{\theta \left( R-r\right) }{R^{3}}+\frac{\theta
\left( r-R\right) }{r^{3}},  \notag \\
&&\tilde{f}_{0i}^{\left( 0\right) }\left( x\right) =B^{\left( 0\right)
i}\left( \mathbf{x}\right) =0\,,  \label{in2B}
\end{eqnarray}%
for its solution. Substituting (\ref{nm8.1.1}) in (\ref{nm8.1}) with $%
\mathcal{F}_{\alpha \beta }(x)=\overline{\mathcal{F}}_{\alpha \beta }$ one
finds that the first-order corrections $f_{\nu \mu }^{\left( 1\right)
}\left( x\right) $ obey the equation%
\begin{eqnarray}
\partial ^{\nu }f_{\nu \mu }^{\left( 1\right) }\left( x\right) &=&\frac{%
\partial }{\partial x_{\tau }}\left[ \frac{1}{2}\left( \mathfrak{L}_{%
\mathfrak{FF}}\overline{\mathcal{F}}_{\alpha \beta }+\mathfrak{L}_{\mathfrak{%
FG}}\overline{\widetilde{\mathcal{F}}}_{\alpha \beta }\right) \overline{%
\mathcal{F}}_{\tau \mu }f^{\left( 0\right) \alpha \beta }\left( x\right) +%
\mathfrak{L}_{\mathfrak{F}}f_{\tau \mu }^{\left( 0\right) }\left( x\right) %
\right]  \notag \\
&+&\frac{\partial }{\partial x_{\tau }}\left[ \frac{1}{2}\left( \mathfrak{L}%
_{\mathfrak{FG}}\overline{\mathcal{F}}_{\alpha \beta }+\mathfrak{L}_{%
\mathfrak{GG}}\overline{\widetilde{\mathcal{F}}}_{\alpha \beta }\right) 
\overline{\widetilde{\mathcal{F}}}_{\tau \mu }f^{\left( 0\right) \alpha
\beta }\left( x\right) +\mathfrak{L}_{\mathfrak{G}}\tilde{f}_{\tau \mu
}^{\left( 0\right) }\left( x\right) \right] \,.  \label{in3}
\end{eqnarray}%
The static electric correction $E^{\left( 1\right) i}\left( \mathbf{x}%
\right) =f_{0i}^{\left( 1\right) }\left( \mathbf{x}\right) $ to the Coulomb
field (\ref{in2}) $E^{\left( 0\right) i}\left( \mathbf{x}\right) $ is
governed by the $\mu =0$ component of equation (\ref{in3}) with all
time derivatives in it set equal to zero%
\begin{equation}
\boldsymbol{\nabla }\cdot \left( \mathbf{E}^{\left( 1\right) }\left( \mathbf{%
x}\right) -\mathbf{\mathcal{E}}\left( \mathbf{x}\right) \right) =0\,,
\label{nnew7a}
\end{equation}%
while the magnetic response $B^{\left( 1\right) i}\left( \mathbf{x}\right)
=-\left( 1/2\right) \varepsilon _{ijk}f^{\left( 1\right) jk}\left( \mathbf{x}%
\right) $ obeys the spacial components of this equation%
\begin{equation}
\boldsymbol{\nabla }\times \left( \mathbf{B}^{\left( 1\right) }\left( 
\mathbf{x}\right) -\mathbf{\mathfrak{H}}\left( \mathbf{x}\right) \right)
=0\,,  \label{nnew7b}
\end{equation}%
where the vectors $\mathbf{\mathcal{E}}\left( \mathbf{x}\right) $ and $%
\mathbf{\mathfrak{H}}\left( \mathbf{x}\right) $ (referred to by us as auxiliary
electric and magnetic fields, respectively), have the form%
\begin{eqnarray}
\mathbf{\mathcal{E}}\left( \mathbf{x}\right) &=&\mathfrak{L}_{\mathfrak{F}}%
\mathbf{E}^{(0)}\left( \mathbf{x}\right)  \notag \\
&-&\left[ \mathfrak{L}_{\mathfrak{FF}}\mathbf{\overline{E}}\cdot \mathbf{E}%
^{(0)}\left( \mathbf{x}\right) +\mathfrak{L}_{\mathfrak{FG}}\mathbf{%
\overline{B}}\cdot \mathbf{E}^{(0)}\left( \mathbf{x}\right) \right] \mathbf{%
\overline{E}}  \notag \\
&-&\left[ \mathfrak{L}_{\mathfrak{FG}}\mathbf{\overline{E}}\cdot \mathbf{E}%
^{(0)}\left( \mathbf{x}\right) +\mathfrak{L}_{\mathfrak{GG}}\mathbf{%
\overline{B}}\cdot \mathbf{E}^{(0)}\left( \mathbf{x}\right) \right] \mathbf{%
\overline{B}}\,,  \label{nnew13a} \\
\mathbf{\mathfrak{H}}\left( \mathbf{x}\right) &=&-\mathfrak{L}_{\mathfrak{G}}%
\mathbf{E}^{(0)}\left( \mathbf{x}\right)  \notag \\
&-&\left[ \mathfrak{L}_{\mathfrak{FF}}\mathbf{\overline{E}}\cdot \mathbf{E}%
^{(0)}\left( \mathbf{x}\right) +\mathfrak{L}_{\mathfrak{FG}}\mathbf{%
\overline{B}}\cdot \mathbf{E}^{(0)}\left( \mathbf{x}\right) \right] \mathbf{%
\overline{B}}  \notag \\
&+&\left[ \mathfrak{L}_{\mathfrak{FG}}\mathbf{\overline{E}}\cdot \mathbf{E}%
^{(0)}\left( \mathbf{x}\right) +\mathfrak{L}_{\mathfrak{GG}}\mathbf{%
\overline{B}}\cdot \mathbf{E}^{(0)}\left( \mathbf{x}\right) \right] \mathbf{%
\overline{E}\,,}  \label{nnew13b}
\end{eqnarray}

The magnetic correction $\mathbf{B}^{\left( 1\right) }\left( \mathbf{x}%
\right) $ is defined by Eq. (\ref{nnew7b}) up to a gradient of a
scalar function $\Omega \left( \mathbf{x}\right) $,%
\begin{equation}
\mathbf{B}^{\left( 1\right) }\left( \mathbf{x}\right) =\mathbf{\mathfrak{H}}%
\left( \mathbf{x}\right) +\boldsymbol{\nabla }\Omega \left( \mathbf{x}%
\right) \,,  \label{sc2.4.1}
\end{equation}%
subject to the condition%
\begin{equation}
\mathbf{\nabla }^{2}\Omega \left( \mathbf{x}\right) =-\boldsymbol{\nabla }%
\cdot \mathbf{\mathfrak{H}}\left( \mathbf{x}\right) \,,  \label{sc2.4.2}
\end{equation}%
as long as $\mathbf{B}^{\left( 1\right) }\left( \mathbf{x}\right) $ must
satisfy Eq. (\ref{bianchi}), $\boldsymbol{\nabla }\cdot \mathbf{B}^{\left(
1\right) }\left( \mathbf{x}\right) =0$. Likewise, the first-order correction
to the electric field $\mathbf{E}^{\left( 1\right) }\left( \mathbf{x}\right) 
$ (\ref{nnew7a}) is defined by $\mathbf{\mathcal{E}}\left( \mathbf{x}\right) 
$ up to the curl of some vector field $\mathbf{W}\left( \mathbf{x}\right) $,%
\begin{equation}
\mathbf{E}^{\left( 1\right) }\left( \mathbf{x}\right) =\mathbf{\mathcal{E}}%
\left( \mathbf{x}\right) +\boldsymbol{\nabla }\times \mathbf{W}\left( 
\mathbf{x}\right) \,,  \label{sc2.4.3}
\end{equation}%
under the condition that $\mathbf{E}^{\left( 1\right) }\left( \mathbf{x}%
\right) $ be consistent with (\ref{bianchi}), $\boldsymbol{\nabla }\times 
\mathbf{E}^{\left( 1\right) }\left( \mathbf{x}\right) =0,$ i. e.,%
\begin{equation*}
-\boldsymbol{\nabla }\times \mathbf{\mathcal{E}}\left( \mathbf{x}\right) =%
\boldsymbol{\nabla }\times \left[ \boldsymbol{\nabla }\times \mathbf{W}%
\left( \mathbf{x}\right) \right] =\boldsymbol{\nabla }\left( \boldsymbol{%
\nabla }\cdot \mathbf{W}\left( \mathbf{x}\right) \right) -\boldsymbol{\nabla 
}^{2}\mathbf{W}\left( \mathbf{x}\right) \,.
\end{equation*}%
By an infinitesimal $U\left( 1\right) $ gauge transformation on the
arbitrary vector field $\mathbf{W}^{\prime }\left( \mathbf{x}\right) =%
\mathbf{W}\left( \mathbf{x}\right) +\boldsymbol{\nabla }\Lambda \left( 
\mathbf{x}\right) $, one may select $\Lambda \left( \mathbf{x}\right) $ to
cancel the term $\boldsymbol{\nabla }\cdot \mathbf{W}\left( \mathbf{x}%
\right) $. Whence%
\begin{equation*}
\boldsymbol{\nabla }\times \mathbf{\mathcal{E}}\left( \mathbf{x}\right) =%
\boldsymbol{\nabla }^{2}\mathbf{W}\left( \mathbf{x}\right) \,,
\end{equation*}%
and one may conclude that both the first-order magnetic field $\mathbf{B}%
^{\left( 1\right) }\left( \mathbf{x}\right) $ and the first-order electric
field $\mathbf{E}^{\left( 1\right) }\left( \mathbf{x}\right) $ are the
transverse and the longitudinal components of $\mathbf{\mathfrak{H}}\left( 
\mathbf{x}\right) $ and $\mathbf{\mathcal{E}}\left( \mathbf{x}\right) $,
respectively%
\begin{equation}
B^{\left( 1\right) i}\left( \mathbf{x}\right) =\left( \delta ^{ij}-\partial
_{i}\frac{1}{\mathbf{\nabla }^{2}}\partial _{j}\right) \mathfrak{H}%
^{j}\left( \mathbf{x}\right) \,,\ \ E^{\left( 1\right) i}\left( \mathbf{x}%
\right) =\partial _{i}\frac{1}{\boldsymbol{\nabla }^{2}}\partial _{j}%
\mathcal{E}^{j}\left( \mathbf{x}\right) \,,  \label{sc2.5}
\end{equation}%
which is written in the integral form as%
\begin{eqnarray}
B^{\left( 1\right) i}\left( \mathbf{x}\right) &=&\mathfrak{H}^{i}\left( 
\mathbf{x}\right) +\frac{1}{4\pi }\partial _{i}\partial _{j}\int d\mathbf{y}%
\frac{\mathfrak{H}^{j}\left( \mathbf{y}\right) }{\left\vert \mathbf{x-y}%
\right\vert }\,,  \label{sc2.5b} \\
E^{\left( 1\right) i}\left( \mathbf{x}\right) &=&-\frac{1}{4\pi }\partial
_{i}\partial _{j}\int d\mathbf{y}\frac{\mathcal{E}^{j}\left( \mathbf{y}%
\right) }{\left\vert \mathbf{x-y}\right\vert }\,.  \label{sc2.5c}
\end{eqnarray}%
In the subsections below we proceed to evaluate the first-order nonlinear
corrections (\ref{sc2.5b}), (\ref{sc2.5c}) owing to an electric charge
distribution (\ref{const charge}), the source of a Coulomb field $\mathbf{E}%
^{\left( 0\right) }\left( \mathbf{x}\right) $ (\ref{in2}), placed in
constant and homogeneous external fields$\mathbf{\mathfrak{\ }\overline{E}}$ and $\mathbf{\overline{B}}$.

\subsection{Electric response\label{ElecC}}

In this subsection we evaluate the first-order electric corrections
following the method in \cite{ShaGit2012}, based on the application of the
projection operator. We leave the detailed consideration of the magnetic
perturbation caused by the same charge distribution (\ref{const charge}) for
a separate article.

\subsubsection{Induced charge density\label{charge}}

Before considering the electric solution (\ref{sc2.5c}) for the field
equation, it is worth highlighting some interesting properties of the
induced anisotropic charge distribution $\rho ^{(1)}\left( \mathbf{x}\right)
=\mathbf{\nabla \cdot E}^{\left( 1\right) }\left( \mathbf{x}\right) =\mathbf{%
\nabla \cdot \mathcal{E}}\left( \mathbf{x}\right) $, as calculated directly
from (\ref{nnew13a}) and (\ref{in2})%
\begin{eqnarray}
\rho ^{(1)}\left( \mathbf{x}\right) &=&\mathfrak{L}_{\mathfrak{F}}\rho
^{(0)}\left( \mathbf{x}\right) -\frac{q}{4\pi }\left( \mathfrak{L}_{%
\mathfrak{FF}}\overline{\mathbf{E}}^{2}+\mathfrak{L}_{\mathfrak{GG}}%
\overline{\mathbf{B}}^{2}-2\mathfrak{L}_{\mathfrak{FG}}\overline{\mathfrak{G}%
}\right) \Phi  \notag \\
&-&\frac{q}{4\pi }\left[ \mathfrak{L}_{\mathfrak{FF}}\left( \mathbf{%
\overline{E}}\cdot \mathbf{x}\right) ^{2}+\mathfrak{L}_{\mathfrak{GG}}\left( 
\mathbf{\overline{B}}\cdot \mathbf{x}\right) ^{2}+2\mathfrak{L}_{\mathfrak{FG%
}}\left( \mathbf{\overline{B}}\cdot \mathbf{x}\right) \left( \mathbf{%
\overline{E}}\cdot \mathbf{x}\right) \right] \frac{\Phi ^{\prime }}{r}\,,
\label{induced dens}
\end{eqnarray}%
where the prime denotes differentiation over $r$ and $\overline{\mathfrak{G}}%
=-\mathbf{\overline{E}}\cdot \mathbf{\overline{B}}$. (As the function $\Phi
(r)$ (\ref{in2}) is continuous at $r=R$ no differentiation of the step
functions in it is needed. This results in the absence of an induced surface
charge.) This density may be presented as%
\begin{equation*}
\rho ^{\left( 1\right) }\left( \mathbf{x}\right) =\rho _{\mathrm{in}%
}^{\left( 1\right) }\left( \mathbf{x}\right) \theta \left( R-r\right) +\rho
_{\mathrm{out}}^{\left( 1\right) }\left( \mathbf{x}\right) \theta \left(
r-R\right) \,,
\end{equation*}%
where%
\begin{eqnarray}
\rho _{\mathrm{in}}^{\left( 1\right) }\left( \mathbf{x}\right) &=&\frac{q}{%
4\pi }\frac{1}{R^{3}}\left[ 3\mathfrak{L}_{\mathfrak{F}}-\left( \mathfrak{L}%
_{\mathfrak{FF}}\overline{\mathbf{E}}^{2}+\mathfrak{L}_{\mathfrak{GG}}%
\overline{\mathbf{B}}^{2}-2\mathfrak{L}_{\mathfrak{FG}}\overline{\mathfrak{G}%
}\right) \right]  \label{rho-in} \\
\rho _{\mathrm{out}}^{\left( 1\right) }\left( \mathbf{x}\right) &=&\frac{q}{%
4\pi }\frac{1}{r^{3}}\left\{ \mathfrak{L}_{\mathfrak{FF}}\left[ 3\left( 
\frac{\mathbf{\overline{E}}\cdot \mathbf{x}}{r}\right) ^{2}-\overline{%
\mathbf{E}}^{2}\right] +\mathfrak{L}_{\mathfrak{GG}}\left[ 3\left( \frac{%
\mathbf{\overline{B}}\cdot \mathbf{x}}{r}\right) ^{2}-\overline{\mathbf{B}}%
^{2}\right] \right.  \notag \\
&+&\left. 2\mathfrak{L}_{\mathfrak{FG}}\left[ 3\left( \frac{\mathbf{%
\overline{B}}\cdot \mathbf{x}}{r}\right) \left( \frac{\mathbf{\overline{E}}%
\cdot \mathbf{x}}{r}\right) +\overline{\mathfrak{G}}\right] \right\} \,.
\label{rho-out}
\end{eqnarray}%
The first term $\mathfrak{L}_{\mathfrak{F}}\rho ^{(0)}\left( \mathbf{x}%
\right) $ in $\rho _{\mathrm{in}}^{(1)}\left( \mathbf{x}\right) $ is
entirely localized inside the sphere $r<R$. On the contrary, the rest part
of the induced density (\ref{induced dens}) is different from zero both
inside and outside that sphere. Nevertheless, the total induced charge $Q_{%
\mathrm{out}}$ found between any two spheres with their radii $r_{1}$ and $r_{2}$
both larger than $R,$ defined as the integral of $\rho ^{(1)}\left( \mathbf{x%
}\right) $ over this volume,$\ $vanishes. To demonstrate this fact note that
the contribution from the term proportional to $\mathfrak{L}_{\mathfrak{FF}}$
at $r>R$ into the volume integral of (\ref{rho-out}) $\overline{\mathbf{E}}%
^{2}-3\left( \frac{\mathbf{\overline{E}}\cdot \mathbf{x}}{r}\right) ^{2}=%
\overline{\mathbf{E}}^{2}(1-3\zeta ^{2}),$ where $\zeta $ is the cosine of
the angle $\theta $ between $\mathbf{\overline{E}}$ and $\mathbf{x}$\textbf{,%
} turns to zero when integrated over $\zeta $ within the limits $\pm 1$. The
same is true for the contribution from the $\mathfrak{L}_{\mathfrak{GG}}$%
 term. It remains to consider the last term in (\ref{rho-out}), the one
proportional to $\mathfrak{L}_{\mathfrak{FG}}$. It is $-\left( \mathbf{%
\overline{E}}\cdot \mathbf{\overline{B}}\right) +3\left( \frac{\mathbf{%
\overline{E}}\cdot \mathbf{x}}{r}\right) \left( \frac{\mathbf{\overline{B}}%
\cdot \mathbf{x}}{r}\right) =-\overline{E}\overline{B}\left( \cos \psi
-3\cos \theta (\cos \psi \cos \theta +\sin \theta \sin \psi \cos \varphi
\right) )$, where $\psi $ is the angle between $\mathbf{\overline{E}}$ and $%
\mathbf{\overline{B}}$\textbf{,} and $\varphi $ is the polar angle. The
integration over $\varphi $ annihilates the last term, and we are left with $%
-\frac{\overline{E}\overline{B}}{r^{3}}\cos \psi (1-3\zeta ^{2})$, which
again disappears after being integrated over $\zeta $. We conclude that%
\begin{equation}
Q_{\mathrm{out}}=0\,.  \label{Q-out}
\end{equation}

The total charge induced inside the sphere may be calculated by integrating (%
\ref{rho-in}) over the sphere to be%
\begin{equation}
Q_{\mathrm{in}}=q\left( \mathfrak{L}_{\mathfrak{F}}+\frac{\tilde{b}}{3}%
\right) \,,  \label{Q-in}
\end{equation}%
where the notation%
\begin{equation}
\tilde{b}=-\left( \mathfrak{L}_{\mathfrak{FF}}\mathbf{\overline{E}}^{2}+%
\mathfrak{L}_{\mathfrak{GG}}\mathbf{\overline{B}}^{2}-2\overline{\mathfrak{G}%
}\mathfrak{L}_{\mathfrak{FG}}\right) \,,  \label{b-tilded}
\end{equation}%
is used. The result (\ref{Q-in}) implies the presence of correction to the
initial charge $q$.\emph{\ }We stress again that Eq. (\ref{b-tilded}), as well
as many other equations above do not have a Lorentz-invariant form, since
they are associated with the special frame where the charge is at rest.

When the core radius shrinks to zero, $R\rightarrow 0,$ the whole charge (%
\ref{Q-in}) is pressed inside, the density $\rho _{\mathrm{in}}^{(1)}\left( 
\mathbf{x}\right) $ tending to infinity. Therefore, finally, the charge
density induced by a point charge is%
\begin{equation}
\rho _{\mathrm{point}}^{(1)}\left( \mathbf{x}\right) =Q_{\mathrm{in}}\delta
^{3}\left( \mathbf{x}\right) +\rho _{\mathrm{out}}^{(1)}\left( \mathbf{x}%
\right) \,.  \label{rho point}
\end{equation}%
A more formal discussion underlying the appearance of the delta function
will be traced in the next item.

\begin{figure}[th!]
\begin{center}
\includegraphics[scale=1.4]{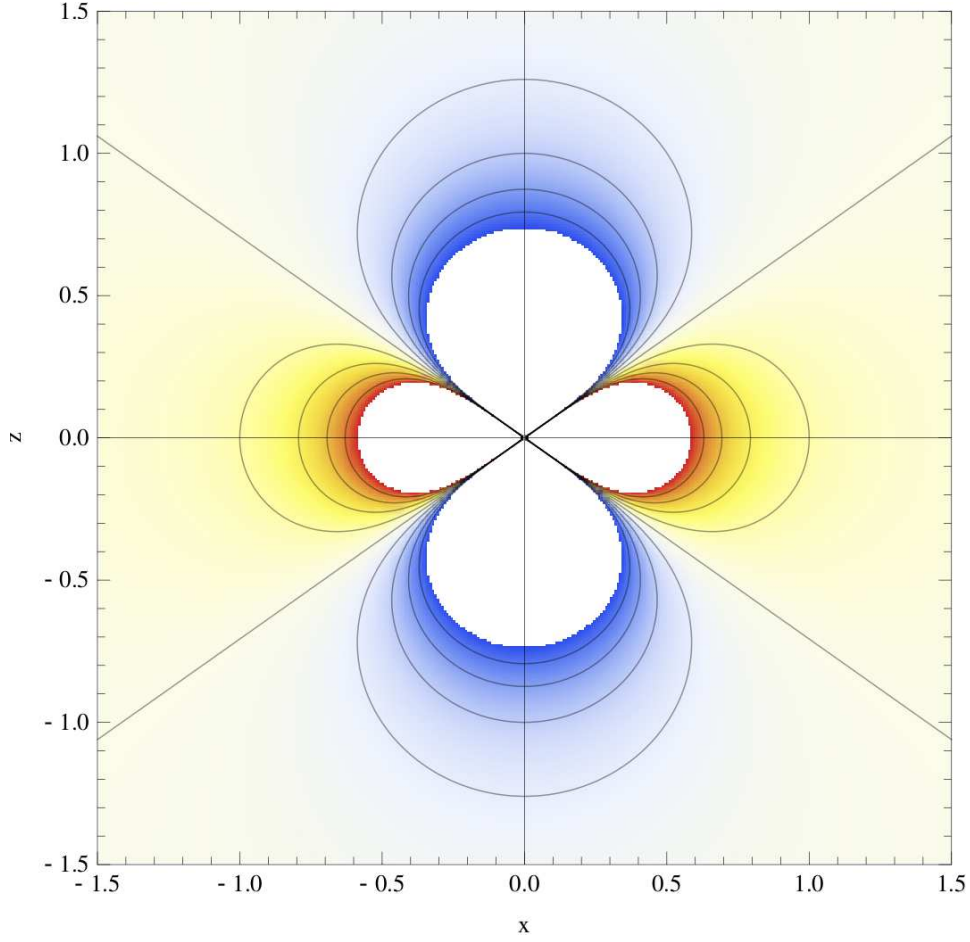}
\end{center}
\caption{The contour plot of the induced charge density (\protect\ref{rho
point2}) in the plane $x_{1}=x\,, x_{3}=z$. The 3-D pattern is to be
obtained by a half-revolution around the axiz $x_{2}=y$. In the polar cones $%
\vert z\vert >\vert x/\protect\sqrt{2}\vert$ $\left(\vert\protect\zeta\vert
<\arccos \left( 1/\protect\sqrt{3}\right)\right)$ -- (blue online) -- the
charge density is negative, while it is positive in the equatorial zone --
(yellow - red online). The solid lines are those of a constant charge
density, the latter being zero on the separating straight lines $z =x/%
\protect\sqrt{2}$. Owing to the $1/r^{3}$ dependence, regions close to the
origin could not be handled by the code and are left blank.}
\label{density-d}
\end{figure}

The outer charge density (\ref{rho-out}) becomes more illustrative for the
special case of the background constant fields parallel to one another, $%
\mathbf{\overline{E}}\parallel \mathbf{\overline{B}}$%
\begin{equation}
\rho _{\mathrm{out}}^{(1)}\left( \mathbf{x}\right) =\frac{q\tilde{b}}{4\pi
r^{3}}(1-3\zeta ^{2})\,,  \label{rho point2}
\end{equation}%
where $\zeta $ is now the cosine of the angle between the radius vector $%
\mathbf{x}$ and the common direction of the fields $\mathbf{\overline{E}}$
and $\mathbf{\overline{B}}$. This form is cylindric symmetrical as
independent of the polar angle. The sign of the charge distributed inside
the two polar cones $1>\zeta ^{2}>\frac{1}{3}$ is opposite to that of the
charge distributed outside them, $\zeta ^{2}<\frac{1}{3}$. So, towards the
common axis of the background fields the induced charge screens locally the
initial charge, whereas in the orthogonal direction the latter is
antiscreened, the global effect of screening by the charge distributed
outside the core being absent, as prescribed by (\ref{Q-out}). The charge
density (\ref{rho point2}) is depicted in Fig. \ref{density-d}.

\subsubsection{Nonlinearly modified Coulomb field\label{field}}

The electric correction (\ref{sc2.5c}) is calculated in Appendix \ref{AppII}%
. It can be split into two parts:%
\begin{equation}
\mathbf{E}^{\left( 1\right) }\left( \mathbf{x}\right) =\mathbf{E}_{\mathrm{in%
}}^{\left( 1\right) }\left( \mathbf{x}\right) \theta \left( R-r\right) +%
\mathbf{E}_{\mathrm{out}}^{\left( 1\right) }\left( \mathbf{x}\right) \theta
\left( r-R\right) \,,  \label{arb14}
\end{equation}%
where the inner $\mathbf{E}_{\mathrm{in}}^{\left( 1\right) }\left( \mathbf{x}%
\right) $ component corresponds to points inside the Coulomb source,%
\begin{eqnarray}
\mathbf{E}_{\mathrm{in}}^{\left( 1\right) }\left( \mathbf{x}\right) &=&\frac{%
q}{4\pi }\frac{1}{R^{3}}\left( \mathfrak{L}_{\mathfrak{F}}+\frac{\tilde{b}}{5%
}\right) \mathbf{x}-\frac{2}{5R^{3}}\left( \frac{q}{4\pi }\right) \left[ 
\mathfrak{L}_{\mathfrak{FF}}\left( \mathbf{\overline{E}}\cdot \mathbf{x}%
\right) +\mathfrak{L}_{\mathfrak{FG}}\left( \mathbf{\overline{B}}\cdot 
\mathbf{x}\right) \right] \mathbf{\overline{E}}  \notag \\
&-&\frac{2}{5R^{3}}\left( \frac{q}{4\pi }\right) \left[ \mathfrak{L}_{%
\mathfrak{FG}}\left( \mathbf{\overline{E}}\cdot \mathbf{x}\right) +\mathfrak{%
L}_{\mathfrak{GG}}\left( \mathbf{\overline{B}}\cdot \mathbf{x}\right) \right]
\mathbf{\overline{B}\,,}  \label{arb14.1}
\end{eqnarray}%
while $\mathbf{E}_{\mathrm{out}}^{\left( 1\right) }\left( \mathbf{x}\right) $
corresponds to the electric field at points outside the Coulomb source%
\begin{eqnarray}
\mathbf{E}_{\mathrm{out}}^{\left( 1\right) }\left( \mathbf{x}\right) &=&%
\frac{q}{4\pi }\left\{ \mathfrak{L}_{\mathfrak{F}}+\left( 1-\frac{3R^{2}}{%
5r^{2}}\right) \frac{\tilde{b}}{2}+\frac{3}{2r^{2}}\left( 1-\frac{R^{2}}{%
r^{2}}\right) \times \right.  \notag \\
&\times &\left. \left[ \mathfrak{L}_{\mathfrak{FF}}\left( \mathbf{\overline{E%
}}\cdot \mathbf{x}\right) ^{2}+\mathfrak{L}_{\mathfrak{GG}}\left( \mathbf{%
\overline{B}}\cdot \mathbf{x}\right) ^{2}+2\mathfrak{L}_{\mathfrak{FG}%
}\left( \mathbf{\overline{B}}\cdot \mathbf{x}\right) \left( \mathbf{%
\overline{E}}\cdot \mathbf{x}\right) \right] \right\} \frac{\mathbf{x}}{r^{3}%
}  \notag \\
&-&\frac{q}{4\pi }\frac{1}{r^{3}}\left( 1-\frac{3R^{2}}{5r^{2}}\right) \left[
\mathfrak{L}_{\mathfrak{FF}}\left( \mathbf{\overline{E}}\cdot \mathbf{x}%
\right) +\mathfrak{L}_{\mathfrak{FG}}\left( \mathbf{\overline{B}}\cdot 
\mathbf{x}\right) \right] \mathbf{\overline{E}}  \notag \\
&-&\frac{q}{4\pi }\frac{1}{r^{3}}\left( 1-\frac{3R^{2}}{5r^{2}}\right) \left[
\mathfrak{L}_{\mathfrak{FG}}\left( \mathbf{\overline{E}}\cdot \mathbf{x}%
\right) +\mathfrak{L}_{\mathfrak{GG}}\left( \mathbf{\overline{B}}\cdot 
\mathbf{x}\right) \right] \mathbf{\overline{B}}\,.  \label{arb14.2}
\end{eqnarray}%
The electric field given by Eqs. (\ref{arb14}), (\ref{arb14.1}), (\ref%
{arb14.2}) is continuous on the surface of the sphere $r=R$, the boundary of
the charge (\ref{const charge}). Note that the $R$-dependent, fast
decreasing at $\left( r/R\right) \rightarrow \infty ,$ part 
\begin{eqnarray*}
\mathbf{E}_{\mathrm{out}}^{\left( 1\right) }\left( \mathbf{x};R\right) &=&%
\frac{q}{4\pi }\left\{ \left( -\frac{3R^{2}}{5r^{2}}\right) \frac{\tilde{b}}{%
2}-\frac{3}{2r^{2}}\left( \frac{R^{2}}{r^{2}}\right) \times \right. \\
&\times &\left. \left[ \mathfrak{L}_{\mathfrak{FF}}\left( \mathbf{\overline{E%
}}\cdot \mathbf{x}\right) ^{2}+\mathfrak{L}_{\mathfrak{GG}}\left( \mathbf{%
\overline{B}}\cdot \mathbf{x}\right) ^{2}+2\mathfrak{L}_{\mathfrak{FG}%
}\left( \mathbf{\overline{B}}\cdot \mathbf{x}\right) \left( \mathbf{%
\overline{E}}\cdot \mathbf{x}\right) \right] \right\} \frac{\mathbf{x}}{r^{3}%
} \\
&+&\frac{q}{4\pi }\frac{1}{r^{3}}\left( \frac{3R^{2}}{5r^{2}}\right) \left[ 
\mathfrak{L}_{\mathfrak{FF}}\left( \mathbf{\overline{E}}\cdot \mathbf{x}%
\right) +\mathfrak{L}_{\mathfrak{FG}}\left( \mathbf{\overline{B}}\cdot 
\mathbf{x}\right) \right] \mathbf{\overline{E}} \\
&+&\frac{q}{4\pi }\frac{1}{r^{3}}\left( \frac{3R^{2}}{5r^{2}}\right) \left[ 
\mathfrak{L}_{\mathfrak{FG}}\left( \mathbf{\overline{E}}\cdot \mathbf{x}%
\right) +\mathfrak{L}_{\mathfrak{GG}}\left( \mathbf{\overline{B}}\cdot 
\mathbf{x}\right) \right] \mathbf{\overline{B}}\,,
\end{eqnarray*}%
of the outer field (\ref{arb14.2}) is a solution of the sourceless equation $%
\mathbf{\nabla \cdot E}_{\mathrm{out}}^{\left( 1\right) }\left( \mathbf{x}%
;R\right) =0$. This is not unexpected, because the source (\ref{rho-out})
does not contain $R$. The hidden role of adding the free solution $\mathbf{E}%
_{\mathrm{out}}^{\left( 1\right) }\left( \mathbf{x};R\right) $ is only to
provide the continuity. It has appeared automatically within the calculation
of the projection (\ref{sc2.5c}).

\subsubsection{Parallel background fields \label{parallel}}

For a simplifying special case of external fields parallel $\mathbf{%
\overline{E}\parallel \overline{B}}$ in the rest frame of the charge (or
antiparallel in the spatially reflected frame)\textbf{,} $\mathbf{\overline{B%
}}=\overline{B}\boldsymbol{\mu }$, $\mu =1$ the results (\ref{arb14.1}) and (%
\ref{arb14.2}) above take the cylindric-symmetrical form $\mathbf{E}%
^{(1)}=\alpha (r,\zeta )\mathbf{x}+\beta (r,\zeta )\boldsymbol{\mu }$, namely%
\begin{equation}
\mathbf{E}_{\mathrm{in}}^{\left( 1\right) }\left( \mathbf{x}\right) =\frac{q%
}{4\pi }\left[ \frac{1}{5R^{3}}\left( 5\mathfrak{L}_{\mathfrak{F}}+\tilde{b}%
\right) \mathbf{x}+\frac{2\tilde{b}r}{5R^{3}}\left( \frac{\boldsymbol{\mu }%
\cdot \mathbf{x}}{r}\right) \boldsymbol{\mu }\right] \,,  \label{arb15.1}
\end{equation}%
and%
\begin{eqnarray}
\mathbf{E}_{\mathrm{out}}^{\left( 1\right) }\left( \mathbf{x}\right)  &=&%
\frac{q}{4\pi }\left[ \mathfrak{L}_{\mathfrak{F}}+\frac{\tilde{b}}{2}\left(
1-\frac{3R^{2}}{5r^{2}}\right) -\frac{3\tilde{b}}{2}\left( 1-\frac{R^{2}}{%
r^{2}}\right) \left( \frac{\boldsymbol{\mu }\cdot \mathbf{x}}{r}\right) ^{2}%
\right] \frac{\mathbf{x}}{r^{3}}  \notag \\
&+&\frac{q}{4\pi }\frac{\tilde{b}}{r^{2}}\left( 1-\frac{3R^{2}}{5r^{2}}%
\right) \left( \frac{\boldsymbol{\mu }\cdot \mathbf{x}}{r}\right) 
\boldsymbol{\mu }\,.  \label{arb15.2}
\end{eqnarray}%
The structural functions $\alpha (r,\zeta )$ and $\beta (r,\zeta )%
\boldsymbol{\ }$are, respectively, even and odd in the angular variable $%
\zeta =\left( \mathbf{x}\cdot \boldsymbol{\mu }\right) /r$ because $%
\boldsymbol{\mu }$ is a pseudovector, while $q\tilde{b}$ and $q\mathfrak{L}_{%
\mathfrak{F}}$ are scalars. In the limit $R/r\rightarrow 0$, the free part%
\begin{equation}
\frac{3\tilde{b}q}{8\pi }R^{2}\left\{ \left[ -\frac{1}{5}+\left( \frac{%
\boldsymbol{\mu }\cdot \mathbf{x}}{r}\right) ^{2}\right] \frac{\mathbf{x}}{%
r^{5}}-\frac{2}{5}\left( \frac{\boldsymbol{\mu }\cdot \mathbf{x}}{r}\right) 
\frac{\boldsymbol{\mu }}{r^{4}}\right\} \,,  \label{free par}
\end{equation}%
of the outer field disappears, such that (\ref{arb15.2}) becomes the
electric field generated by a pointlike particle $\mathbf{E}_{\mathrm{PL}%
}^{\left( 1\right) }\left( \mathbf{x}\right) $, valid for every $r>0$%
\begin{equation}
\lim_{R\rightarrow 0}\mathbf{E}_{\mathrm{out}}^{\left( 1\right) }\left( 
\mathbf{x}\right) =\mathbf{E}_{\mathrm{PL}}^{\left( 1\right) }\left( \mathbf{%
x}\right) =\frac{q}{4\pi }\left[ \mathfrak{L}_{\mathfrak{F}}+\frac{\tilde{b}%
}{2}\left( 1-3\zeta ^{2}\right) \right] \frac{\mathbf{x}}{r^{3}}+\frac{q}{%
4\pi }\frac{\tilde{b}}{r^{2}}\zeta \boldsymbol{\mu }\,,\ \ r>0\,,
\label{r>0}
\end{equation}%
with $\zeta =\frac{\boldsymbol{\mu }\cdot \mathbf{x}}{r}$ being the azimuth
angle cosine. The induced charge density corresponding to such a limit $\rho
_{\mathrm{PL}}^{\left( 1\right) }=\boldsymbol{\nabla }\cdot \mathbf{E}_{%
\mathrm{PL}}^{\left( 1\right) }\left( \mathbf{x}\right) $ distributed
outside the origin is%
\begin{equation*}
\rho _{\mathrm{PL}}^{\left( 1\right) }=\frac{q\tilde{b}}{4\pi r^{3}}\left(
1-3\zeta ^{2}\right) \,,\text{ }r>0\,.
\end{equation*}%
Therefore, the total charge concentrated in the volume $V$ between any two
spheres of finite radii $r_{1}$ and $r_{2}$ is zero:%
\begin{equation*}
\int_{V}\rho _{\mathrm{PL}}^{\left( 1\right) }d^{3}x=\frac{q\tilde{b}}{2}%
\int_{r_{1}}^{r_{2}}\frac{dr}{r}\int_{-1}^{1}\left( 1-3\zeta ^{2}\right)
d\zeta =0\,,
\end{equation*}%
in agreement with the general case of nonparallel background fields (\ref%
{Q-out}). On the other hand, the flux of the field (\ref{r>0}) through the
surface of a sphere centered in the origin is different from zero (and
independent of the radius)%
\begin{equation}
\oint_{S}\mathbf{E}_{\mathrm{PL}}^{\left( 1\right) }\left( \mathbf{x}\right)
\cdot \mathbf{\hat{n}}dS=q\left( \mathfrak{L}_{\mathfrak{F}}+\frac{1}{3}%
\tilde{b}\right) \,,  \label{cor-charge}
\end{equation}%
which coincides with (\ref{Q-in}). Because of the Gauss theorem $\oint_{S}%
\mathbf{E}^{\left( 1\right) }\left( \mathbf{x}\right) \cdot \mathbf{\hat{n}}%
dS=\int_{V}d^{3}x\mathbf{\nabla \cdot E}^{\left( 1\right) }\left( \mathbf{x}%
\right) $ this implies that the distribution of the induced charge $\rho
^{(1)}=\mathbf{\nabla \cdot E}^{\left( 1\right) }\left( \mathbf{x}\right) $
in parallel constant background fields is%
\begin{equation}
\rho _{\mathrm{point}}^{(1)}=\frac{q}{4\pi }\left[ \left( \mathfrak{L}_{%
\mathfrak{F}}+\frac{\tilde{b}}{3}\right) \delta ^{3}\left( \mathbf{x}\right)
+\frac{\tilde{b}}{r^{3}}\left( 1-3\left( \frac{\boldsymbol{\mu }\cdot 
\mathbf{x}}{r}\right) ^{2}\right) \right] \,,  \notag
\end{equation}%
following the definition of the Dirac delta-function in terms of an
integral. Note that formally the second term here, as well as in (\ref{rho
point}), has a $r^{-3}$ singularity in the origin. Certainly this
singularity should not be taken seriously, because the whole infrared
approximation on which our present approach is based fails in this point.
Nevertheless this singularity itself is not significant, since the overall
distributed charge is zero thanks to the angle integration.

\subsubsection{Scalar potential \label{potential}}

As long as the fulfillment of identities (\ref{bianchi}) is provided, the
result for the modified Coulomb field must have a potential representation $%
\mathbf{E}^{\left( 1\right) }\left( \mathbf{x}\right) =-\boldsymbol{\nabla }%
A_{0}^{(1)}\left( \mathbf{x}\right) $. For the cylindrically symmetric case
of parallel background fields the field (\ref{arb15.1}) and (\ref{arb15.2})
corresponds to the potential%
\begin{equation*}
A_{0}^{(1)}\left( \mathbf{x}\right) =A_{0}^{(1)\mathrm{in}}\left( \mathbf{x}%
\right) \theta \left( R-r\right) +A_{0}^{(1)\mathrm{out}}\left( \mathbf{x}%
\right) \theta \left( r-R\right) \,,
\end{equation*}%
where its inner and outer parts are of the form%
\begin{equation*}
A_{0}^{(1)\mathrm{in}}\left( \mathbf{x}\right) =\mathrm{const.}+r^{2}\lambda
_{1}\left( \zeta ^{2}\right) \,,\text{ \ }A_{0}^{(1)\mathrm{out}}\left( 
\mathbf{x}\right) =\frac{\lambda _{2}\left( \zeta ^{2}\right) }{r}+\frac{%
\lambda _{3}\left( \zeta ^{2}\right) }{r^{3}}\,,
\end{equation*}%
namely%
\begin{eqnarray}
A_{0}^{(1)\mathrm{in}}\left( \mathbf{x}\right) &=&\frac{q}{4\pi }\left\{ 
\frac{1}{2R}\left( 3\mathfrak{L}_{\mathfrak{F}}+\tilde{b}\right) -\frac{1}{%
5R^{3}}\left[ \frac{1}{2}\left( \tilde{b}+5\mathfrak{L}_{\mathfrak{F}%
}\right) +\tilde{b}\zeta ^{2}\right] r^{2}\right\} \,,  \notag \\
A_{0}^{(1)\mathrm{out}}\left( \mathbf{x}\right) &=&\frac{q}{4\pi }\left\{ 
\frac{1}{r}\left[ \mathfrak{L}_{\mathfrak{F}}+\frac{\tilde{b}}{2}\left(
1-\zeta ^{2}\right) \right] +\frac{\tilde{b}R^{2}}{10r^{3}}\left( 3\zeta
^{2}-1\right) \right\} \,.  \label{A-out}
\end{eqnarray}%
The structural coefficient functions $\lambda _{1,2,3}\left( \zeta
^{2}\right) $ are even in $\zeta =\frac{\mathbf{x\cdot }\overline{\mathbf{B}}%
}{r\overline{B}}$ because the latter is a pseudoscalar, while $\tilde{b}$
and $\mathfrak{L}_{\mathfrak{F}}$ are scalars. The written potential is
continuous at $r=R$ everywhere at the sphere surface. Its part $\frac{q}{%
4\pi }\frac{\tilde{b}R^{2}}{10r^{3}}\left( 3\zeta ^{2}-1\right) $ depending
on $R$ satisfies the free equation $\boldsymbol{\nabla }^{2}A_{0}^{(1)%
\mathrm{out}}=0$ out of the core, as it should. This is the quadrupole
addition to the potential in the case of cylindric symmetry \cite%
{LanLifFieldTheor}, the quadrupole moment being $\frac{q}{4\pi }\frac{\tilde{%
b}R^{2}}{5}$. It disappears together with $R$, which means that for the
pointlike initial charge the induced charge contribution does not possess a
quadrupole moment. The boundary conditions for the scalar potential have
been chosen so as to provide finiteness at $r=0$ and decreasing at $%
r\rightarrow \infty $.

We present in Fig. 2 the pattern of electric lines of force following Eq.
(42) and the equipotential lines, corresponding to the pointlike limit of
Eq. (\ref{A-out}) at $r>0$%
\begin{equation}
A_{0}^{(1)\mathrm{PL}}\left( \mathbf{x}\right) =\frac{q}{4\pi }\frac{1}{r}%
\left[ \mathfrak{L}_{\mathfrak{F}}+\frac{\tilde{b}}{2}\left( 1-\zeta
^{2}\right) \right] \,,  \label{A-out-PL}
\end{equation}%
which is also the asymptotic behavior of the potential Eq. (\ref{A-out}) in
the far-off domain.

\begin{figure}[th]
\begin{center}
\includegraphics[scale=0.92]{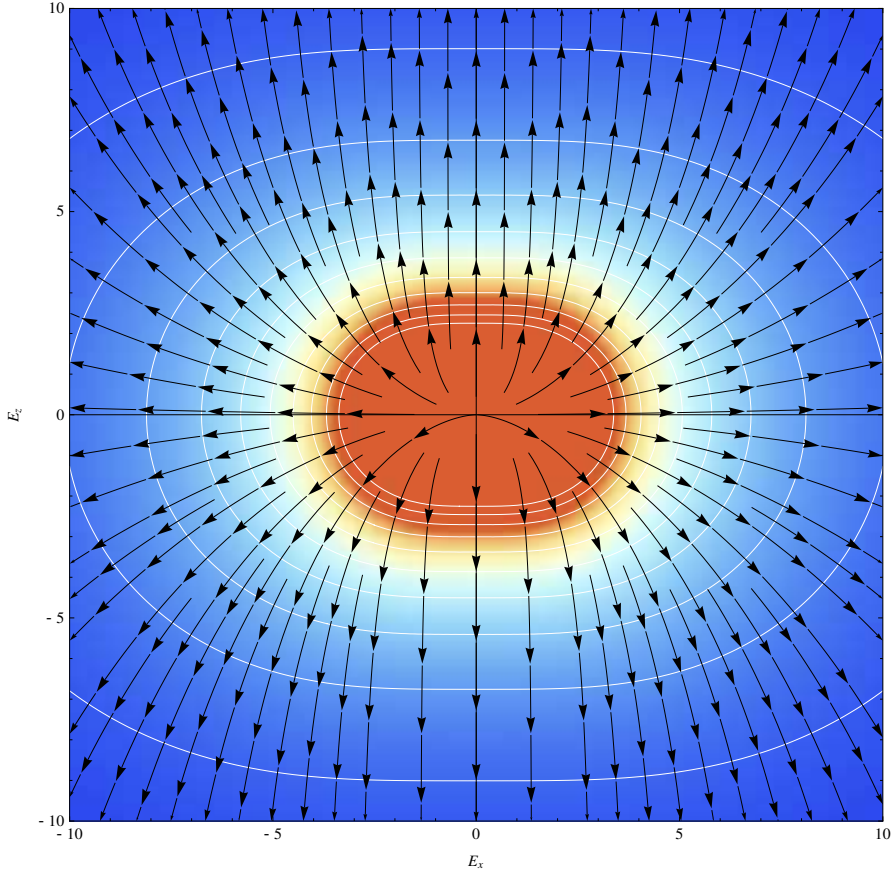}
\end{center}
\caption{Electric lines of force (\protect\ref{r>0}) (represented by black
arrows) and the corresponding equipotential curves (\protect\ref{A-out-PL})
(represented by white lines) of a point-like source placed in external
constant electric and magnetic fields directed along the $z$ axis. We set $%
q>0$ and $\left( q/4\protect\pi \right) \mathfrak{L}_{\mathfrak{F}}=1=\left(
q/4\protect\pi \right) \tilde{b}$. The electric field is more intense close
to the origin, represented by the red region, becoming weaker far from the
origin, represented by the blue regions. The complete pattern corresponds to
the modification of the standard Coulombic electric lines of force and its
equipotential lines.}
\label{PL-electric-2}
\end{figure}
It remains to confront the present results with those obtained previously
for purely magnetic background \cite{shabus} \cite{SadSod07}. For that case,
using a different method, the following expression for the scalar potential
of a point charge in a magnetic field $\overline{\mathbf{B}}$ in the far-off
region (the anisotropic Coulomb law) was finally presented as Eq. (27) in Ref.
\cite{ShaGit2012}:%
\begin{equation}
A_{0}(\mathbf{x})=\frac{q}{4\pi }\frac{1}{\sqrt{\varepsilon _{\perp }}\sqrt{%
\varepsilon _{\perp }x_{_{\parallel }}^{2}+\varepsilon _{_{\parallel }}%
\mathbf{x}_{\perp }^{2}}},  \label{coulomb}
\end{equation}%
where $\varepsilon _{\perp }=1-\mathfrak{L}_{\mathfrak{F}}$ and \ $%
\varepsilon _{_{\parallel }}=1-\mathfrak{L}_{\mathfrak{F}}+2\overline{%
\mathfrak{F}}\mathfrak{L}_{\mathfrak{G}\mathfrak{G}}\ $are\ eigenvalues of
the dielectric tensor \cite{villalbachaves} $\varepsilon _{ij}=\left( 1-%
\mathfrak{L}_{\mathfrak{F}}\right) \delta _{ij}+\mathfrak{L}_{\mathfrak{G}%
\mathfrak{G}}\overline{B}_{i}\overline{B}_{j},$ responsible for polarization
induced by homogeneously charged planes parallel and orthogonal to $%
\overline{\mathbf{B}}$, respectively, \cite{Shabus2011}. In Eq. (\ref%
{coulomb}), $\mathbf{x}_{\perp }=r\sin \theta $ and $\mathbf{x}_{\parallel
}=r\cos \theta $ are the coordinate components across and along $\overline{%
\mathbf{B}}$, respectively. Taking this into account we rewrite the anisotropic Coulomb
potential (\ref{coulomb}) as%
\begin{equation}
A_{0}(\mathbf{x})=\frac{q}{4\pi r}\frac{1}{\sqrt{\varepsilon _{\perp }}\sqrt{%
\varepsilon _{\perp }\cos ^{2}\theta +\varepsilon _{_{\parallel }}\sin
^{2}\theta }}=\frac{q}{4\pi r}\frac{1}{\sqrt{\varepsilon _{\perp }}\sqrt{1-%
\mathfrak{L}_{\mathfrak{F}}+2\overline{\mathfrak{F}}\mathfrak{L}_{\mathfrak{G%
}\mathfrak{G}}\sin ^{2}\theta }}.  \label{coulomb2}
\end{equation}%
Considering the corrections $\mathfrak{L}_{\mathfrak{F}}$ and $2\overline{%
\mathfrak{F}}\mathfrak{L}_{\mathfrak{G}\mathfrak{G}}$ to the vacuum
dielectric permeability $\epsilon =1$ as small, we find that in this
approximation Eq. (\ref{coulomb2}) turns into $\frac{q}{4\pi r}\left[ 1+%
\mathfrak{L}_{\mathfrak{F}}-\mathfrak{L}_{\mathfrak{GG}}\overline{\mathfrak{F%
}}\left( 1-\zeta ^{2}\right) \right] ,$ which is just $\frac{q}{4\pi r}%
+A_{0}^{(1)\mathrm{PL}}\left( \mathbf{x}\right) ,$ Eq. (\ref{A-out-PL}),
with $\tilde{b}=-2\mathfrak{L}_{\mathfrak{GG}}\overline{\mathfrak{F}}$
obtained by setting $\mathbf{\overline{E}}^{2}=\overline{\mathfrak{G}}=0$ in
(\ref{b-tilded}). The important difference between the results (\ref{coulomb}%
) or (\ref{coulomb2}), on one hand, and (\ref{A-out-PL}), on the other, is
that the former were obtained using the full photon propagator component $%
D_{00}=$ ($\boldsymbol{k}^{2}-\varkappa _{2})^{-1}$ (in the momentum
representation) with $\varkappa _{2}=\boldsymbol{k}^{2}\mathfrak{L}_{%
\mathfrak{F}}-k_{\parallel }^{2}2\overline{\mathfrak{F}}\mathfrak{L_{GG}}$.
Here $\boldsymbol{k}^{2}=\boldsymbol{k}_{\perp }^{2}+k_{\parallel }^{2},$
and $\boldsymbol{k}_{\perp }$ and $k_{\parallel }$ are the momentum components
across and along $\overline{\boldsymbol{B}}$ , respectively. The full
propagator is decomposed into the series 
\begin{equation}
D_{00}=\frac{1}{\boldsymbol{k}^{2}-\varkappa _{2}}=\frac{1}{\boldsymbol{k}%
^{2}}+\frac{1}{\boldsymbol{k}^{2}}\frac{\varkappa _{2}}{\boldsymbol{k}^{2}}%
+...,  \label{D}
\end{equation}%
corresponding to the summation of the one-photon-reducible chain of one-
electron-positron loop diagrams. On the contrary, the correction (\ref%
{A-out-PL}) corresponds to the second term in this expansion alone, since
only the free photon propagator $\frac{1}{\boldsymbol{\nabla }^{2}}=\frac{1}{%
\boldsymbol{k}^{2}}$ was used in (\ref{sc2.5}) for finding the field of a
point charge. As a consequence, there is a difference in the convexity of
the lines of force in Fig. 2 as compared to Fig. 2b in Ref. \cite{13th
Lomonosov} and, correspondingly, in the orientation of the equipotential
ellipsoid as compared to Fig. 1 in Ref. \cite{GodVys}. The summation of the
chain, though not strictly grounded, was crucial for establishing \cite%
{shabus} the squeezing of the Coulomb field into a string in the limit of
infinite magnetic field and for the screening of the Coulomb field by a
strong magnetic field that prevents the collapse of the hydrogen atom in the
limit $\overline{B}\rightarrow \infty $ (see also \cite{SadSod07, GodVys}).

\section{Considerations within the Euler-Heisenberg effective Lagrangian 
\label{EH}}

It is most interesting to apply the formulas of the previous sections valid
for the arbitrary local nonlinear theory to quantum electrodynamics approximated
by the Euler-Heisenberg (EH) effective Lagrangian \cite{EH}. This is what we
shall be doing in the present Section where, for the sake of simplicity, we
omit bars over the background fields.

According to Refs. \cite{EH,BerLifPit} it is written as%
\begin{equation}
\mathfrak{L}=\frac{m^{4}}{8\pi ^{2}}\int_{0}^{\infty }dt\frac{e^{-t}}{t^{3}}%
\left\{ -\left( ta\cot ta\right) \left( tb\coth tb\right) +1-\frac{1}{3}%
\left( a^{2}-b^{2}\right) t^{2}\right\} \,,  \label{eh1}
\end{equation}%
where the integration contour is meant to circumvent the poles on the real
axis of $t$ supplied by $\cot ta$ above the real axis. Because of the
exponentially fast decrease of the integrand at $t\rightarrow \infty $ we
may equivalently accept that the integration path is a straight ray inclined
by an infinitesimally small angle to the real axis in the upper complex
plane. This circumstance will be significant in the situation treated in
Sec. \ref{Strong} where the Schwinger effect of pair creation from
the vacuum might come into play.

In (\ref{eh1}), the dimensionless invariant combinations, eigenvalues of the
field tensor,%
\begin{eqnarray}
&&a=\left( \frac{e}{m^{2}}\right) \sqrt{-\mathfrak{F}+\sqrt{\mathfrak{F}^{2}+%
\mathfrak{G}^{2}}}\,,  \notag \\
&&b=\left( \frac{e}{m^{2}}\right) \sqrt{\mathfrak{F}+\sqrt{\mathfrak{F}^{2}+%
\mathfrak{G}^{2}}}\,  \label{eh2}
\end{eqnarray}%
have the meaning of the electric and magnetic field in a Lorentz frame in
which these are parallel, normalized to the characteristic field value $%
\frac{e}{m^{2}}$, where $m$ and $e$ are the electron mass and charge, respectively. Such a
frame exists as long as $\mathfrak{G}\neq 0.$ Since in the previous sections
we mostly dealt with the general case where the fields are not necessarily
parallel in the rest frame of the charge, we shall not generally identify $a$
with $eE/m^{2}$ and $b$ with $eB/m^{2}$.

\subsection{Small background fields}

In the small-field limit, $a,b\ll 1,$ we should take the (lowest) quadratic
approximation for the Euler-Heisenberg Lagrangian (\ref{eh1}) known to be%
\begin{equation*}
\mathfrak{L}_{\mathrm{weak}}\simeq \frac{1}{8\pi ^{2}}\left( \frac{e^{4}}{%
m^{4}}\right) \left( \frac{4\mathfrak{F}^{2}+7\mathfrak{G}^{2}}{45}\right)
\,.
\end{equation*}%
Hence, the coefficients involved in our previous formulas should be defined
in this approximation by Eqs. (\ref{nm8.0}) with $\mathcal{F}=0$. One has $%
\mathfrak{L}_{\mathfrak{FF}}=\frac{1}{8\pi ^{2}}\left( \frac{e^{4}}{m^{4}}%
\right) \left( \frac{8}{45}\right) ,$ $\mathfrak{L}_{\mathfrak{GG}}=\frac{1}{%
8\pi ^{2}}\left( \frac{e^{4}}{m^{4}}\right) \left( \frac{14}{45}\right) ,$ $%
\mathfrak{L}_{\mathfrak{FG}}=0$, and $\mathfrak{L}_{\mathfrak{F}}=0$. With this
substitution, the induced charge Eq. (\ref{Q-in}), for instance becomes%
\begin{equation*}
Q_{\mathrm{in}}=-\frac{q}{12\pi ^{2}}\left( \frac{e^{4}}{45m^{4}}\right)
\left( 4\mathbf{E}^{2}+7\mathbf{B}^{2}\right) \,.
\end{equation*}%
\ 

\subsection{Strong background fields}

In the general case of arbitrarily strong fields the necessary coefficients
are calculated from (\ref{ec1}). With the help of (\ref{eh2}) and two
auxiliary functions $\mathcal{H}\left( \tau \right) $ and $\mathcal{Q}\left(
\tau \right) $,%
\begin{equation}
\mathcal{H}\left( \tau \right) =\tau \coth \tau -\frac{\tau ^{2}}{\sinh
^{2}\tau }\,,\ \ \mathcal{Q}\left( \tau \right) =\tau \cot \tau -\frac{\tau
^{2}}{\sin ^{2}\tau }\,,  \label{eh3}
\end{equation}%
the derivative of (\ref{eh1}) with respect to $\mathfrak{F}$ can be
expressed as%
\begin{eqnarray}
\mathfrak{L}_{\mathfrak{F}} &=&\frac{m^{4}}{16\pi ^{2}}\frac{b^{2}}{\sqrt{%
\mathfrak{F}^{2}+\mathfrak{G}^{2}}}\int_{0}^{\infty }d\tau \frac{e^{-\tau /b}%
}{\tau ^{3}}\left\{ -\mathcal{H}\left( \tau \right) \left( \frac{a\tau }{b}%
\cot \frac{a\tau }{b}\right) \right.   \notag \\
&+&\left. \left( \tau \coth \tau \right) \mathcal{Q}\left( \frac{a\tau }{b}%
\right) +\frac{2\tau ^{2}}{3}\left[ 1+\left( \frac{a}{b}\right) ^{2}\right]
\right\} \,.  \label{eh4}
\end{eqnarray}%
Introducing two more constants $\gamma _{\pm }$ and new auxiliary functions $%
\mathcal{\tilde{H}}\left( \tau \right) $ and $\mathcal{\tilde{Q}}\left( \tau
\right) $,%
\begin{eqnarray}
&&\gamma _{\pm }=1\pm \frac{2\mathfrak{F}}{\sqrt{\mathfrak{F}^{2}+\mathfrak{G%
}^{2}}}\,,  \notag \\
&&\mathcal{\tilde{Q}}\left( \tau \right) =\frac{2\tau ^{2}}{\sin ^{2}\tau }%
\left( \tau \cot \tau -1\right) \,,\ \ \mathcal{\tilde{H}}\left( \tau
\right) =\frac{2\tau ^{2}}{\sinh ^{2}\tau }\left( \tau \coth \tau -1\right)
\,.  \label{eh4.2}
\end{eqnarray}%
the second-order derivatives involved in the definition of $\tilde{b}$ (\ref%
{b-tilded}) are:%
\begin{eqnarray}
\mathfrak{L}_{\mathfrak{FF}} &=&\frac{m^{4}}{32\pi ^{2}}\frac{b^{2}}{\left( 
\mathfrak{F}^{2}+\mathfrak{G}^{2}\right) }\int_{0}^{\infty }d\tau \frac{%
e^{-\tau /b}}{\tau ^{3}}\left\{ 2\mathcal{H}\left( \tau \right) \mathcal{Q}%
\left( \frac{a\tau }{b}\right) \right.   \notag \\
&-&\tau \coth \tau \left[ \gamma _{+}\mathcal{Q}\left( \frac{a\tau }{b}%
\right) +\mathcal{\tilde{Q}}\left( \frac{a\tau }{b}\right) \right]   \notag
\\
&-&\left. \left( \frac{a\tau }{b}\cot \frac{a\tau }{b}\right) \left[ \gamma
_{-}\mathcal{H}\left( \tau \right) +\mathcal{\tilde{H}}\left( \tau \right) %
\right] \right\} \,,  \notag \\
\mathfrak{L}_{\mathfrak{GG}} &=&\frac{m^{4}}{32\pi ^{2}}\frac{b^{2}}{\left( 
\mathfrak{F}^{2}+\mathfrak{G}^{2}\right) }\int_{0}^{\infty }d\tau \frac{%
e^{-\tau /b}}{\tau ^{3}}\left\{ -2\mathcal{H}\left( \tau \right) \mathcal{Q}%
\left( \frac{a\tau }{b}\right) \right.   \notag \\
&-&\tau \coth \tau \left[ -\gamma _{+}\mathcal{Q}\left( \frac{a\tau }{b}%
\right) +\frac{b^{2}}{a^{2}}\mathcal{\tilde{Q}}\left( \frac{a\tau }{b}%
\right) \right]   \notag \\
&-&\left. \left( \frac{a\tau }{b}\cot \frac{a\tau }{b}\right) \left[ -\gamma
_{-}\mathcal{H}\left( \tau \right) +\frac{a^{2}}{b^{2}}\mathcal{\tilde{H}}%
\left( \tau \right) \right] \right\} \,,  \notag \\
\mathfrak{L}_{\mathfrak{FG}} &=&\frac{m^{4}\kappa }{32\pi ^{2}}\frac{b^{2}}{%
\left( \mathfrak{F}^{2}+\mathfrak{G}^{2}\right) }\int_{0}^{\infty }d\tau 
\frac{e^{-\tau /b}}{\tau ^{3}}\left\{ \frac{1}{ab}\left( a^{2}-b^{2}\right) 
\mathcal{H}\left( \tau \right) \mathcal{Q}\left( \frac{a\tau }{b}\right)
\right.   \notag \\
&+&\frac{b\tau }{a}\coth \tau \left[ -\gamma _{-}\mathcal{Q}\left( \frac{%
a\tau }{b}\right) +\mathcal{\tilde{Q}}\left( \frac{a\tau }{b}\right) \right] 
\notag \\
&+&\left. \frac{a}{b}\left( \frac{a\tau }{b}\cot \frac{a\tau }{b}\right) %
\left[ \gamma _{+}\mathcal{H}\left( \tau \right) -\mathcal{\tilde{H}}\left(
\tau \right) \right] \right\} \,.  \label{eh5}
\end{eqnarray}%
It should be noted that $\mathfrak{L}_{\mathfrak{FG}}$ as a pseudoscalar is
an odd function, as expressed by the constant $\kappa =\mathrm{sign}\left( 
\mathfrak{G}\right) $.

\subsubsection{Magnetic-dominated large-field regime}

We can represent Eqs. (\ref{eh4}), (\ref{eh5}) as functions of $b$ and of
the ratio $a/b$, that can be then expanded in powers of the latter assuming
that 
\begin{equation}
\frac{a}{b}\ll 1,  \label{eh6}
\end{equation}%
but irrespective of whether $a\ $\ and $b$ are small or not as compared to
unity. Condition (\ref{eh6}) implies the magnetic dominance $B\gg E$ in any
reference frame. The expansion of (\ref{eh4}) and (\ref{eh5}) includes the
expansion of the trigonometric cotangent in the power series of its
argument. Thereby the poles at the real axis are avoided in accord with the
well-known fact that an expansion over the (relatively) small electric field
excludes the Schwinger effect. Then by virtue of (\ref{eh6}), every
coefficient is expressed as a power series of the ratio $a/b$,%
\begin{equation}
\mathfrak{L}_{k}=\mathfrak{L}_{k}^{\left( 0\right) }+\left( \frac{a}{b}%
\right) ^{2}\mathfrak{L}_{k}^{\left( 2\right) }+O\left[ \left( \frac{a}{b}%
\right) ^{4}\right] \,,  \label{expan}
\end{equation}%
where $k$ simply labels distinct coefficients, i. e., $k=\left\{ \mathfrak{F}%
,\mathfrak{FF},\mathfrak{GG},\mathfrak{FG}\right\} $. It should be noted
that the identity $\mathfrak{L}_{\mathfrak{FG}}^{\left( 0\right) }=0$ is a
consequence of the fact that $\mathfrak{L}_{\mathfrak{FG}}$ must be absent
when the electric field is zero. We obtain, for the leading $\mathfrak{L}%
_{k}^{\left( 0\right) }$ and the next-to-leading $\mathfrak{L}_{k}^{\left(
2\right) }$ terms, the following expressions%
\begin{eqnarray}
\mathfrak{L}_{\mathfrak{F}}^{\left( 0\right) } &=&\frac{\alpha }{2\pi }%
\int_{0}^{\infty }d\tau \frac{e^{-\tau /b}}{\tau }\left( \frac{2}{3}-\frac{%
\coth \tau }{\tau }+\frac{1}{\sinh ^{2}\tau }\right) \,,  \notag \\
\mathfrak{L}_{\mathfrak{F}}^{\left( 2\right) } &=&\frac{\alpha }{2\pi }%
\int_{0}^{\infty }d\tau \frac{e^{-\tau /b}}{\tau }\left[ \left( \frac{1}{%
\tau }-\frac{\tau }{3}\right) \coth \tau -\left( 1+\frac{\tau ^{2}}{3}%
\right) \frac{1}{\sinh ^{2}\tau }\right] \,,  \label{eh24.8}
\end{eqnarray}%
for the first derivatives and%
\begin{eqnarray}
\left( \mathfrak{F}+\sqrt{\mathfrak{F}^{2}+\mathfrak{G}^{2}}\right) 
\mathfrak{L}_{\mathfrak{FF}}^{\left( 0\right) } &=&\frac{\alpha }{2\pi }%
\int_{0}^{\infty }d\tau \frac{e^{-\tau /b}}{\tau }\left( \frac{1}{\tau }%
\coth \tau +\frac{1}{\sinh ^{2}\tau }-\frac{2\tau \coth \tau }{\sinh
^{2}\tau }\right) \,,  \notag \\
\left( \mathfrak{F}+\sqrt{\mathfrak{F}^{2}+\mathfrak{G}^{2}}\right) 
\mathfrak{L}_{\mathfrak{FF}}^{\left( 2\right) } &=&\frac{\alpha }{2\pi }%
\int_{0}^{\infty }d\tau \frac{e^{-\tau /b}}{\tau }\left[ \left( \tau -\frac{6%
}{\tau }\right) \coth \tau +\frac{\left( 2+\tau ^{2}\right) }{\sinh ^{2}\tau 
}\right.   \notag \\
&+&\left. \left( \frac{2\tau ^{2}}{3}+4\right) \frac{\tau \coth \tau }{\sinh
^{2}\tau }\right] \,,  \notag \\
\left( \mathfrak{F}+\sqrt{\mathfrak{F}^{2}+\mathfrak{G}^{2}}\right) 
\mathfrak{L}_{\mathfrak{GG}}^{\left( 0\right) } &=&\frac{\alpha }{2\pi }%
\int_{0}^{\infty }d\tau \frac{e^{-\tau /b}}{\tau }\left( \frac{2\tau }{3}%
\coth \tau -\frac{1}{\tau }\coth \tau +\frac{1}{\sinh ^{2}\tau }\right) \,, 
\notag \\
\left( \mathfrak{F}+\sqrt{\mathfrak{F}^{2}+\mathfrak{G}^{2}}\right) 
\mathfrak{L}_{\mathfrak{GG}}^{\left( 2\right) } &=&\frac{\alpha }{2\pi }%
\int_{0}^{\infty }d\tau \frac{e^{-\tau /b}}{\tau }\left[ \left( \frac{4\tau
^{3}}{15}+\frac{6}{\tau }-\frac{5\tau }{3}\right) \coth \tau \right.   \notag
\\
&-&\left. \left( 4+\frac{5\tau ^{2}}{3}\right) \frac{1}{\sinh ^{2}\tau }-%
\frac{2\tau \coth \tau }{\sinh ^{2}\tau }\right] \,,  \notag \\
\left( \mathfrak{GL}_{\mathfrak{FG}}\right) ^{\left( 2\right) } &=&\frac{%
\alpha }{2\pi }\int_{0}^{\infty }d\tau \frac{e^{-\tau /b}}{\tau }\left[
\left( \frac{3}{\tau }-\frac{2\tau }{3}\right) \coth \tau \right.   \notag \\
&-&\left. \left( 1+\frac{2\tau ^{2}}{3}\right) \frac{1}{\sinh ^{2}\tau }-%
\frac{2\tau \coth \tau }{\sinh ^{2}\tau }\right]\,,   \label{eh24.9}
\end{eqnarray}%
for the second derivatives.

This representation allows us to consider the asymptotic regime of large
magnetic field with the electric field kept moderate by studying the $b\gg 1$
limit. In this regime the coefficient\ $\mathfrak{L}_{\mathfrak{GG}}$
dominates over every other coefficient in our expressions for the fields and
charges, since the integrals (\ref{eh24.8}) and (\ref{eh24.9}) over $\tau $
would diverge linearly for $\left(\mathfrak{F}+\sqrt{\mathfrak{F}^{2}+\mathfrak{G}^{2}} \right)\mathfrak{L}_{\mathfrak{GG}}^{\left( 0\right) }$
and cubically for $\left(\mathfrak{F}+\sqrt{\mathfrak{F}^{2}+\mathfrak{G}^{2}} \right)\mathfrak{L}_{\mathfrak{GG}}^{\left( 2\right) }$ if the
limit $b=\infty $ had been formally substituted into their integrands. To be
more rigorous, the asymptotic forms of Eqs. (\ref{eh24.8}) and (\ref{eh24.9})
read%
\begin{equation}
\mathfrak{L}_{\mathfrak{F}}^{\left( 0\right) }\simeq \frac{\alpha }{2\pi }%
\left( \frac{2}{3}\log b+\mathcal{K}_{\mathfrak{F}}^{\left( 0\right)
}\right) \,,\ \ \mathfrak{L}_{\mathfrak{F}}^{\left( 2\right) }\simeq \frac{%
\alpha }{2\pi }\left( -\frac{b}{3}+\mathcal{K}_{\mathfrak{F}}^{\left(
2\right) }\right) \,,\ \ \left( \mathcal{K}_{\mathfrak{F}}^{\left( 0\right)
},\mathcal{K}_{\mathfrak{F}}^{\left( 2\right) }\right) =\mathrm{const.}\,,
\label{LFasy}
\end{equation}%
and%
\begin{eqnarray}
&&\left( \mathfrak{F}+\sqrt{\mathfrak{F}^{2}+\mathfrak{G}^{2}}\right) 
\mathfrak{L}_{\mathfrak{FF}}^{\left( 0\right) }\simeq \frac{\alpha }{2\pi }%
\mathcal{K}_{\mathfrak{FF}}^{\left( 0\right) }\,,\ \ \mathcal{K}_{\mathfrak{%
FF}}^{\left( 0\right) }=\mathrm{const.\,,}  \notag \\
&&\left( \mathfrak{F}+\sqrt{\mathfrak{F}^{2}+\mathfrak{G}^{2}}\right) 
\mathfrak{L}_{\mathfrak{FF}}^{\left( 2\right) }\simeq \frac{\alpha }{2\pi }%
\left( b+\mathcal{K}_{\mathfrak{FF}}^{\left( 2\right) }\right) \,,\ \ 
\mathcal{K}_{\mathfrak{FF}}^{\left( 2\right) }=\mathrm{const.\,,}  \notag \\
&&\left( \mathfrak{F}+\sqrt{\mathfrak{F}^{2}+\mathfrak{G}^{2}}\right) 
\mathfrak{L}_{\mathfrak{GG}}^{\left( 0\right) }\simeq \frac{\alpha }{2\pi }%
\left( \frac{2}{3}b+\mathcal{K}_{\mathfrak{GG}}^{\left( 0\right) }\right)
\,,\ \ \mathcal{K}_{\mathfrak{GG}}^{\left( 0\right) }=\mathrm{const.\,,} 
\notag \\
&&\left( \mathfrak{F}+\sqrt{\mathfrak{F}^{2}+\mathfrak{G}^{2}}\right) 
\mathfrak{L}_{\mathfrak{GG}}^{\left( 2\right) }\simeq \frac{\alpha }{2\pi }%
\left( \frac{8}{15}b^{3}-\frac{5}{3}b+\mathcal{K}_{\mathfrak{GG}}^{\left(
2\right) }\right) \,,\ \ \mathcal{K}_{\mathfrak{GG}}^{\left( 2\right) }=%
\mathrm{const.\,,}  \notag \\
&&\left( \mathfrak{GL}_{\mathfrak{FG}}\right) ^{\left( 2\right) }\simeq 
\frac{\alpha }{2\pi }\left( -\frac{2}{3}b+\mathcal{K}_{\mathfrak{FG}%
}^{\left( 2\right) }\right) \,,\ \ \mathcal{K}_{\mathfrak{FG}}^{\left(
2\right) }=\mathrm{const.\,.}  \label{LGGasy}
\end{eqnarray}%
From these results, we find, for instance, that the correction to the
Coulomb charge (\ref{Q-in}) in the magnetic-field-dominant asymptotic regime
limit $\left( a/b\right) \ll 1$, $b\gg 1$ has the form%
\begin{equation}
Q_{\mathrm{strong}}\simeq q\left( \frac{\alpha }{3\pi }\right) \left[ -\frac{%
b}{3}\left( 1+\frac{4}{5}a^{2}\right) +\log b+\mathcal{K}%
\right] \,,\ \ \mathcal{K}\simeq -1.156\,.  \label{eh14}
\end{equation}%
(Here we have expressed $E$ and $B$ in terms of the Lorentz invariants,
although the form (\ref{eh14}) is applicable only in the rest frame of the
charge).

The negative sign in the charge correction here may be attributed to the
lack of asymptotic freedom in QED. Note that in a non-Abelian situation, when
the asymptotic freedom takes place, the linear growth with the magnetic
field is absent from the polarization operator, while the logarithmic term
has an opposite sign (see the discussion around Eqs. (78) - (80) in Ref. \cite%
{Shabus2011}). As applied to effective Lagrangian, connection between the
asymptotic freedom and large-field behavior was first discussed in Ref. \cite%
{Ritus2loop}.

We are now in a position to shed light upon whether the linear growth of the
polarization operator with the magnetic field established earlier in the
pure magnetic case \cite{Skobelev, Shabad book} and resulting \cite{shabus}
in strong screening of the Coulomb field by the magnetic field and
prevention of the collapse of the hydrogen atom \cite{shabus,SadSod07,13th
Lomonosov,MacVys11,GodVys} retains when an electric field is also present.

This problem is burdened by the fact that, in contrast to the pure magnetic
case where only the virtual photons of the eigenmode labeled as mode 2 in Ref.
\cite{BatSha1971, Shabad book} are carriers of electrostatic field \cite%
{Shabus2009}, in that case also mode-3 photons contribute into it. However,
it can be seen from the analysis in Ref. \cite{ShaUs2010} where the eigenvalue
problem of the polarization operator was considered that within the infrared
approximation, in the magnetic-dominating regime, like the one now under
consideration, the above statement reestablishes\footnote{%
This occurs owing to the asymptotic domination of $b$ in accord with $%
\mathfrak{L}_{\mathfrak{GG}}$ in (\ref{eh24.9}). As a result, the functions $%
\Lambda _{2,3}$ in Eqs. (18), (20) of Ref. \cite{ShaUs2010} cannot compete
with $\Lambda _{4}$ (19) in forming the polarization eigenvectors, Eq. (15)
of that reference. This indicates that the scalar potential $\flat _{0}^{(2)}
$ in mode 2 is nonzero in the static limit, while that in mode 3, $\flat
_{0}^{(3)},$ diappears according to Eqs.(39) in \cite{ShaUs2010}, which, in
its turn, follows from Eq. (11) of the same Ref. in the magnetic-domination
regime.}.

Therefore, again the only contribution of the second mode into the photon
propagator (\ref{D}) is responsible for the Coulomb field modification.
According to Eq. (12) of Ref. \cite{ShaUs2010} the polarization operator
eigenvalue $\varkappa _{2}$ is%
\begin{equation*}
\varkappa _{2}=\Lambda _{4}=-\mathbf{k}^{2}\mathfrak{L}_{\mathfrak{F}%
}+k_{3}^{2}\tilde{b}\,,
\end{equation*}%
where we set $k_{0}=0$, and $\mathbf{k}$ and $k_{3}$ are the momentum
components across and along the dominant magnetic field direction. When used
in the photon propagator it corrects the Coulomb potential as%
\begin{equation*}
A_{0}(\mathbf{x})=\frac{q}{4\pi }\int \frac{d^{3}k\exp (-i\mathbf{k\cdot x)}%
}{\mathbf{k}^{2}+\varkappa _{2}}\,.
\end{equation*}%
This means that the polarization correction found in the present paper, when
summed up to the denominator as in (\ref{D}), leads to suppression of the
Coulomb field with the growth of the magnetic field in the same way as it
did when the electric field was absent. This result cannot be immediately
applied to the hydrogen atom, because the direct interaction of the electric
field with the orbital electron in strong magnetic field needs to be taken
into account at a step, previous to considering the radiative corrections
contained in the effective Lagrangian.

\subsubsection{Equally strong magnetic and electric fields\label{Strong}}

Another interesting-to-treat option is the case where both fields are
sufficiently strong, with equal amplitude $a=b\equiv c$, such that the field
invariant $\mathfrak{F}$ vanishes,%
\begin{equation}
\mathfrak{F}=0\,,\ \ \left( \frac{m^{2}}{e}\right) ^{2}c^{2}=\left\vert 
\mathfrak{G}\right\vert \,.  \label{ec1}
\end{equation}%
In this case the coefficient (\ref{eh4}) $\mathfrak{L}_{\mathfrak{F}}$ takes
the form,%
\begin{equation}
\mathfrak{L}_{\mathfrak{F}}=\frac{\alpha }{4\pi }\int_{0}^{\infty }d\tau
e^{-\tau /c}\left( \frac{\cot \tau }{\sinh ^{2}\tau }-\frac{\coth \tau }{%
\sin ^{2}\tau }+\frac{4}{3\tau }\right) \,,  \label{ec2}
\end{equation}%
while $\mathfrak{L}_{\mathfrak{FF}}$, $\mathfrak{L}_{\mathfrak{GG}}$ and $%
\mathfrak{L}_{\mathfrak{FG}}$ reduce to:%
\begin{eqnarray}
\mathfrak{L}_{\mathfrak{FF}} &=&\frac{m^{4}}{32\pi ^{2}}\frac{c^{2}}{%
\mathfrak{G}^{2}}\int_{0}^{\infty }d\tau \frac{e^{-\tau /c}}{\tau ^{3}}%
\left\{ 2\mathcal{H}\left( \tau \right) \mathcal{Q}\left( \tau \right)
\right.   \notag \\
&-&\left. \tau \coth \tau \left( \mathcal{Q}\left( \tau \right) +\mathcal{%
\tilde{Q}}\left( \tau \right) \right) -\tau \cot \tau \left( \mathcal{H}%
\left( \tau \right) +\mathcal{\tilde{H}}\left( \tau \right) \right) \right\}
\,,  \notag \\
\mathfrak{L}_{\mathfrak{GG}} &=&\frac{m^{4}}{32\pi ^{2}}\frac{c^{2}}{%
\mathfrak{G}^{2}}\int_{0}^{\infty }d\tau \frac{e^{-\tau /c}}{\tau ^{3}}%
\left\{ -2\mathcal{H}\left( \tau \right) \mathcal{Q}\left( \tau \right)
\right.   \notag \\
&-&\left. \tau \coth \tau \left[ -\mathcal{Q}\left( \tau \right) +\mathcal{%
\tilde{Q}}\left( \tau \right) \right] -\tau \cot \tau \left[ -\mathcal{H}%
\left( \tau \right) +\mathcal{\tilde{H}}\left( \tau \right) \right] \right\}
\,,  \notag \\
\mathfrak{L}_{\mathfrak{FG}} &=&\kappa \frac{\alpha }{8\pi }\left( \frac{e}{%
m^{2}}\right) ^{2}\frac{1}{c^{2}}\int_{0}^{\infty }d\tau \frac{e^{-\tau /c}}{%
\tau ^{3}}\left\{ \tau \coth \tau \left( -\mathcal{Q}\left( \tau \right) +%
\mathcal{\tilde{Q}}\left( \tau \right) \right) \right.   \notag \\
&+&\left. \tau \cot \tau \left[ \mathcal{H}\left( \tau \right) -\mathcal{%
\tilde{H}}\left( \tau \right) \right] \right\} \,.  \label{ec3}
\end{eqnarray}%
Referring to the choice of the integration path indicated after Eq. (\ref{EH}%
) we see that all trigonometric functions in denominators of (\ref{ec2}), (%
\ref{ec3}) grow fast in the remote integration domain. This observation
allows us to consider the leading asymptotic behavior of these integrals in
the same way as in the previous Subsubsection despite the Schwinger effect.
The integrals (\ref{ec2}), (\ref{ec3}) with $c$ set equal to $\infty $ in
them diverge at the most logarithmically. More rigorous consideration
confirms that the leading terms are (at most) logarithmic and real, i.e.
independent of the Schwinger effect. The absence of the linear growth
manifests that contrary to the case of the previous Subsubsection there is
no strong suppression of the electrostatic field by a constant external
field when both the electric and magnetic components of the latter are
equally strong.

\section{Conclusions}

We addressed the charge distribution induced in the vacuum by the applied
field of an electric charge of finite size $R$ in the background of constant
and homogeneous electric and magnetic fields, and also the electric field
strength and its scalar potential. The field of the charge was treated as
small, but the background was taken into account exactly. So, although the
response to the charge was taken as linear, the charge and the Coulomb field
modification induced by it (even in the limit of small background) are at
least quadratic in the strength of the backgrounds, the overall power of
nonlinearity handled being at least cubic, as seen in (\ref{rho-in}), (\ref%
{rho-out}). Of course, the same problem might have been considered with
better precision, especially in what concerns small distances from the
charge when its size tends to zero, using the available expression for the
polarization operator in the external constant and homogeneous field \cite%
{BatSha1971} (see also \cite{Shabad book}) along the lines of Ref. \cite%
{ShaUs2010}. However, the use of the local (infrared) approximation as
described in \cite{ShaGit2012} allows obtaining much more transparent
results, quite reliable unless $R$ is too small.

We have found that the induced anisotropic charge density is distributed
both within the site of the imposed charge and outside it. The outer part of
the distribution does not depend on the size $R$ of the imposed charge,
while the inner part does, and it tends to delta function when the size
shrinks to zero $R\rightarrow 0$. The distributions occupying the northern
and southern cones of the outer space contain the charges with the sign
opposite to the one occupying the rest of the space. When integrated over
the spherical angle the charge cancels so that any finite outer volume
between two spheres remains neutral. Thus, the net charge is nonzero and it
is located inside the inner volume. In this point the situation drastically
differs from the one claimed for the radiative correction for the Coulomb
charge without background, where the correction to the core (point) charge
compensates the induced distributed charge -- see \cite{MilStra1983} for
(3+1)-dimensional QED and \cite{Khalilov} for (2+1)-dimensional theory as
applied to graphene. The reason lies in the fact that the correction $Q_{%
\mathrm{in}}$ Eq. (\ref{Q-in}), to the charge is reduced to the effect of a
constant dielectric permeability, which cannot help being unity when the
background is absent or disappears far from the charge: the coefficient $%
\tilde{b}$ (\ref{b-tilded}) nullifies in the absence of background $\mathbf{%
\overline{E}=\overline{B}=0,}$ and so does $\mathfrak{L}_{\mathfrak{F}},$
because the correspondence principle requires that there be no quadratic
correction to the Lagrangian for small fields. In the present case, when the
background is nonvanishing in the infinitely remote region, so is the
dielectric constant. It may be thought that the compensating charge $-Q_{%
\mathrm{in}}$ is moved to infinity or concentrates at the outer edge of the
background field, if this edge is imagined to exist -- in full analogy with the
electrodynamics of the media.

The induced distributed charge density $\rho _{\mathrm{out}}$ (\ref{rho-out}%
) decays as $r^{-3}$\ far from the charge. The modified Coulomb field (\ref%
{arb14.1}), (\ref{arb14.2})\ is anisotropic and continuous at the border of
the imposed charge, which implies the absence of the surface charge at this
border. The field depends on the size $R$, the part independent of $R$
decreases as $r^{-2}$, while the $R$-depending part behaves like an electric
quadrupole. At last, but not least, we analyze the nonlinear effects by
considering the one-loop effective action of QED in constant backgrounds,
provided by Euler and Heisenberg. The results shows that a nontrivial
electric field superposed with a constant magnetic field enhances the
screening of the Coulomb field, when compared with the case of a pure
magnetic field \cite{shabus,SadSod07,13th Lomonosov,MacVys11,GodVys}.

\section*{Acknowledgements}

The authors thank the support of the Russian Science Foundation, Research
Project No. 15-12-10009.

\begin{appendices}

\section{Nonlinear Maxwell equations truncated to the cubic power}\label{Ap1}

In this Appendix we present expansions of the coefficient tensors $\frac{%
\delta \mathfrak{L}\left( \mathfrak{F},\mathfrak{G}\right) }{\delta 
\mathfrak{F}\left( x\right) }F_{\tau \mu }\left( x\right) $ and $\frac{%
\delta \mathfrak{L}\left( \mathfrak{F},\mathfrak{G}\right) }{\delta 
\mathfrak{G}\left( x\right) }\tilde{F}_{\tau \mu }\left( x\right) $ involved
in (\ref{MaxEq}) in powers of deviations $f^{\alpha \beta }\left( x\right)
=F^{\alpha \beta }\left( x\right) -\mathcal{F}^{\alpha \beta }\left(
x\right) $ from the background $F^{\alpha \beta }\left( x\right) =\mathcal{F}%
^{\alpha \beta }\left( x\right) $. Although only time- and space-independent
background fields are efficiently handled in the article, all the equations
below are valid irrespective of the condition $\mathcal{F}^{\alpha \beta
}\left( x\right) =\mathrm{Const.}$, i. e. for arbitrary background produced
by the current $\mathcal{J}_{\mu }\left( x\right) $ via Eq. (\ref{background}%
). For the sake of convenience, representing the coefficient tensors as%
\begin{eqnarray}
\frac{\delta \mathfrak{L}\left( \mathfrak{F},\mathfrak{G}\right) }{\delta 
\mathfrak{F}\left( z\right) }F_{\tau \mu }\left( z\right) &=&\lambda _{\tau
\mu }\left( z\right) =\lambda \left( z\right) F_{\tau \mu }\left( z\right)
\,,\ \ \lambda \left( z\right) =\frac{\delta \mathfrak{L}\left( \mathfrak{F},%
\mathfrak{G}\right) }{\delta \mathfrak{F}\left( z\right) }\,,  \notag \\
\frac{\delta \mathfrak{L}\left( \mathfrak{F},\mathfrak{G}\right) }{\delta 
\mathfrak{G}\left( z\right) }\tilde{F}_{\tau \mu }\left( z\right) &=&\tilde{%
\lambda}_{\tau \mu }\left( z\right) =\tilde{\lambda}\left( z\right) \tilde{F}%
_{\tau \mu }\left( z\right) \,,\ \ \tilde{\lambda}\left( z\right) =\frac{%
\delta \mathfrak{L}\left( \mathfrak{F},\mathfrak{G}\right) }{\delta 
\mathfrak{G}\left( z\right) }\,,  \label{nm4new}
\end{eqnarray}%
the expansions reads%
\begin{eqnarray}
\lambda _{\tau \mu }\left( z\right) &=&\left. \lambda _{\tau \mu }\left(
z\right) \right\vert _{f=0}+\int dy\left( \frac{\delta \lambda _{\tau \mu
}\left( z\right) }{\delta F^{\alpha \beta }\left( y\right) }\right)
_{f=0}f^{\alpha \beta }\left( y\right)  \notag \\
&+&\frac{1}{2}\int dydy^{\prime }\left( \frac{\delta ^{2}\lambda _{\tau \mu
}\left( z\right) }{\delta F^{\alpha \beta }\left( y\right) \delta F^{\rho
\sigma }\left( y^{\prime }\right) }\right) _{f=0}f^{\alpha \beta }\left(
y\right) f^{\rho \sigma }\left( y^{\prime }\right)  \notag \\
&+&\frac{1}{6}\int dydy^{\prime }dy^{\prime \prime }\left( \frac{\delta
^{3}\lambda _{\tau \mu }\left( z\right) }{\delta F^{\alpha \beta }\left(
y\right) \delta F^{\rho \sigma }\left( y^{\prime }\right) \delta
F^{\varkappa \gamma }\left( y^{\prime \prime }\right) }\right)
_{f=0}f^{\alpha \beta }\left( y\right) f^{\rho \sigma }\left( y^{\prime
}\right) f^{\varkappa \gamma }\left( y^{\prime \prime }\right) +O\left(
f^{4}\right) \,,  \label{nm5.1}
\end{eqnarray}%
and%
\begin{eqnarray}
\tilde{\lambda}_{\tau \mu }\left( z\right) &=&\left. \tilde{\lambda}_{\tau
\mu }\left( z\right) \right\vert _{f=0}+\int dy\left( \frac{\delta \tilde{%
\lambda}_{\tau \mu }\left( z\right) }{\delta F^{\alpha \beta }\left(
y\right) }\right) _{f=0}f^{\alpha \beta }\left( y\right)  \notag \\
&+&\frac{1}{2}\int dydy^{\prime }\left( \frac{\delta ^{2}\tilde{\lambda}%
_{\tau \mu }\left( z\right) }{\delta F^{\alpha \beta }\left( y\right) \delta
F^{\rho \sigma }\left( y^{\prime }\right) }\right) _{f=0}f^{\alpha \beta
}\left( y\right) f^{\rho \sigma }\left( y^{\prime }\right)  \notag \\
&+&\frac{1}{6}\int dydy^{\prime }dy^{\prime \prime }\left( \frac{\delta ^{3}%
\tilde{\lambda}_{\tau \mu }\left( z\right) }{\delta F^{\alpha \beta }\left(
y\right) \delta F^{\rho \sigma }\left( y^{\prime }\right) \delta
F^{\varkappa \gamma }\left( y^{\prime \prime }\right) }\right)
_{f=0}f^{\alpha \beta }\left( y\right) f^{\rho \sigma }\left( y^{\prime
}\right) f^{\varkappa \gamma }\left( y^{\prime \prime }\right) +O\left(
f^{4}\right) \,.  \label{nm5.2}
\end{eqnarray}

Truncating the series above at a given power of the field deviations $f^{\mu
\nu }\left( z\right) $, say $n$-th power, generates a set of nonlinear
Maxwell equations whose solutions are interpreted as the $n$-th
electromagnetic response of the background field $\mathcal{F}^{\alpha \beta
}\left( x\right) $ to a small electromagnetic source, denoted by $%
j^{\mu }\left( x\right) $ in agreement with the division $J_{\mu }\left(
x\right) =\mathcal{J}_{\mu }\left( x\right) +j_{\mu }\left( x\right) $. In
this Appendix we present an explicit derivation of each coefficient above up
to the third power in the field deviations, whose solutions correspond to a
cubic response of the background to the source $j^{\mu }\left(
x\right) $. We list below each functional derivatives of (\ref{nm5.1}) and (%
\ref{nm5.2}).

\begin{itemize}
\item First functional derivatives:%
\begin{eqnarray}
&&\frac{\delta \lambda _{\tau \mu }\left( z\right) }{\delta F^{\alpha \beta
}\left( y\right) }=\frac{\delta \lambda \left( z\right) }{\delta F^{\alpha
\beta }\left( y\right) }F_{\tau \mu }\left( z\right) +\frac{1}{2}\lambda
\left( z\right) \left( \eta _{\alpha \tau }\eta _{\beta \mu }-\eta _{\alpha
\mu }\eta _{\beta \tau }\right) \delta \left( z-y\right) \,,  \notag \\
&&\frac{\delta \tilde{\lambda}_{\tau \mu }\left( z\right) }{\delta F^{\alpha
\beta }\left( y\right) }=\frac{\delta \tilde{\lambda}\left( z\right) }{%
\delta F^{\alpha \beta }\left( y\right) }\tilde{F}_{\tau \mu }\left(
z\right) +\frac{1}{2}\tilde{\lambda}\left( z\right) \varepsilon _{\tau \mu
\alpha \beta }\delta \left( z-y\right) \,;  \label{nm5.3}
\end{eqnarray}

\item Second functional derivatives:%
\begin{eqnarray}
\frac{\delta ^{2}\lambda _{\tau \mu }\left( z\right) }{\delta F^{\alpha
\beta }\left( y\right) \delta F^{\rho \sigma }\left( y^{\prime }\right) } &=&%
\frac{\delta ^{2}\lambda \left( z\right) }{\delta F^{\alpha \beta }\left(
y\right) \delta F^{\rho \sigma }\left( y^{\prime }\right) }F_{\tau \mu
}\left( z\right) +\frac{1}{2}\frac{\delta \lambda \left( z\right) }{\delta
F^{\alpha \beta }\left( y\right) }\left( \eta _{\tau \rho }\eta _{\mu \sigma
}-\eta _{\tau \sigma }\eta _{\mu \rho }\right) \delta \left( z-y^{\prime
}\right)  \notag \\
&+&\frac{1}{2}\frac{\delta \lambda \left( z\right) }{\delta F^{\rho \sigma
}\left( y^{\prime }\right) }\left( \eta _{\tau \alpha }\eta _{\mu \beta
}-\eta _{\tau \beta }\eta _{\mu \alpha }\right) \delta \left( z-y\right) \,,
\notag \\
\frac{\delta ^{2}\tilde{\lambda}_{\tau \mu }\left( z\right) }{\delta
F^{\alpha \beta }\left( y\right) \delta F^{\rho \sigma }\left( y^{\prime
}\right) } &=&\frac{\delta ^{2}\tilde{\lambda}\left( z\right) }{\delta
F^{\alpha \beta }\left( y\right) \delta F^{\rho \sigma }\left( y^{\prime
}\right) }\tilde{F}_{\tau \mu }\left( z\right) +\frac{1}{2}\frac{\delta 
\tilde{\lambda}\left( z\right) }{\delta F^{\alpha \beta }\left( y\right) }%
\varepsilon _{\tau \mu \rho \sigma }\delta \left( z-y^{\prime }\right) 
\notag \\
&+&\frac{1}{2}\frac{\delta \tilde{\lambda}\left( z\right) }{\delta F^{\rho
\sigma }\left( y^{\prime }\right) }\varepsilon _{\tau \mu \alpha \beta
}\delta \left( z-y\right) \,;  \label{nm5.4}
\end{eqnarray}

\item Third functional derivatives:%
\begin{eqnarray}
\frac{\delta ^{3}\lambda _{\tau \mu }\left( z\right) }{\delta F^{\alpha
\beta }\left( y\right) \delta F^{\rho \sigma }\left( y^{\prime }\right)
\delta F^{\varkappa \gamma }\left( y^{\prime \prime }\right) } &=&\frac{%
\delta ^{3}\lambda \left( z\right) }{\delta F^{\alpha \beta }\left( y\right)
\delta F^{\rho \sigma }\left( y^{\prime }\right) \delta F^{\varkappa \gamma
}\left( y^{\prime \prime }\right) }F_{\tau \mu }\left( z\right)  \notag \\
&+&\frac{1}{2}\frac{\delta ^{2}\lambda \left( z\right) }{\delta F^{\alpha
\beta }\left( y\right) \delta F^{\rho \sigma }\left( y^{\prime }\right) }%
\left( \eta _{\gamma \mu }\eta _{\varkappa \tau }-\eta _{\gamma \tau }\eta
_{\varkappa \mu }\right) \delta \left( z-y^{\prime \prime }\right)  \notag \\
&+&\frac{1}{2}\frac{\delta ^{2}\lambda \left( z\right) }{\delta F^{\alpha
\beta }\left( y\right) \delta F^{\varkappa \gamma }\left( y^{\prime \prime
}\right) }\left( \eta _{\tau \rho }\eta _{\mu \sigma }-\eta _{\tau \sigma
}\eta _{\mu \rho }\right) \delta \left( z-y^{\prime }\right)  \notag \\
&+&\frac{1}{2}\frac{\delta ^{2}\lambda \left( z\right) }{\delta F^{\rho
\sigma }\left( y^{\prime }\right) \delta F^{\varkappa \gamma }\left(
y^{\prime \prime }\right) }\left( \eta _{\tau \alpha }\eta _{\mu \beta
}-\eta _{\tau \beta }\eta _{\mu \alpha }\right) \delta \left( z-y\right) \,,
\label{nm5.5}
\end{eqnarray}%
and%
\begin{eqnarray}
\frac{\delta ^{3}\tilde{\lambda}_{\tau \mu }\left( z\right) }{\delta
F^{\alpha \beta }\left( y\right) \delta F^{\rho \sigma }\left( y^{\prime
}\right) \delta F^{\varkappa \gamma }\left( y^{\prime \prime }\right) } &=&%
\frac{\delta ^{3}\tilde{\lambda}\left( z\right) }{\delta F^{\alpha \beta
}\left( y\right) \delta F^{\rho \sigma }\left( y^{\prime }\right) \delta
F^{\varkappa \gamma }\left( y^{\prime \prime }\right) }\tilde{F}_{\tau \mu
}\left( z\right)  \notag \\
&+&\frac{1}{2}\frac{\delta ^{2}\tilde{\lambda}\left( z\right) }{\delta
F^{\alpha \beta }\left( y\right) \delta F^{\rho \sigma }\left( y^{\prime
}\right) }\varepsilon _{\varkappa \gamma \tau \mu }\delta \left( z-y^{\prime
\prime }\right)  \notag \\
&+&\frac{1}{2}\frac{\delta ^{2}\tilde{\lambda}\left( z\right) }{\delta
F^{\alpha \beta }\left( y\right) \delta F^{\varkappa \gamma }\left(
y^{\prime \prime }\right) }\varepsilon _{\tau \mu \rho \sigma }\delta \left(
z-y^{\prime }\right)  \notag \\
&+&\frac{1}{2}\frac{\delta ^{2}\tilde{\lambda}\left( z\right) }{\delta
F^{\rho \sigma }\left( y^{\prime }\right) \delta F^{\varkappa \gamma }\left(
y^{\prime \prime }\right) }\varepsilon _{\tau \mu \alpha \beta }\delta
\left( z-y\right) \,.  \label{nm5.6}
\end{eqnarray}
\end{itemize}

As long as the effective Lagrangian $\mathfrak{L}$ is a function of the
field invariants $\mathfrak{F}\left( z\right) $, $\mathfrak{G}\left(
z\right) $ only, the functional derivatives of $\lambda \left( z\right) $
takes the form\footnote{%
The functional derivatives of $\tilde{\lambda}\left( z\right) $ can be
obtained from (\ref{nm6.1}), (\ref{nm6.2}) and (\ref{nm6.4}) by the formal
substitution $\lambda \left( z\right) \rightarrow \tilde{\lambda}\left(
z\right) $.}

\begin{itemize}
\item First functional derivatives:%
\begin{equation}
\frac{\delta \lambda \left( z\right) }{\delta F^{\alpha \beta }\left(
y\right) }=\frac{1}{2}\left[ \frac{\partial \lambda \left( z\right) }{%
\partial \mathfrak{F}\left( z\right) }F_{\alpha \beta }\left( z\right) +%
\frac{\partial \lambda \left( z\right) }{\partial \mathfrak{G}\left(
z\right) }\tilde{F}_{\alpha \beta }\left( z\right) \right] \delta \left(
y-z\right) \,;  \label{nm6.1}
\end{equation}

\item Second functional derivatives:%
\begin{eqnarray}
\frac{\delta ^{2}\lambda \left( z\right) }{\delta F^{\alpha \beta }\left(
y\right) \delta F^{\rho \sigma }\left( y^{\prime }\right) } &=&\frac{1}{4}%
\left\{ \left( \frac{\partial ^{2}\lambda \left( z\right) }{\partial 
\mathfrak{F}\left( z\right) \partial \mathfrak{G}\left( z\right) }\tilde{F}%
_{\rho \sigma }\left( z\right) +\frac{\partial ^{2}\lambda \left( z\right) }{%
\partial \mathfrak{F}\left( z\right) ^{2}}F_{\rho \sigma }\left( z\right)
\right) F_{\alpha \beta }\left( z\right) \right.  \notag \\
&+&\left( \frac{\partial ^{2}\lambda \left( z\right) }{\partial \mathfrak{G}%
\left( z\right) \partial \mathfrak{F}\left( z\right) }F_{\rho \sigma }\left(
z\right) +\frac{\partial ^{2}\lambda \left( z\right) }{\partial \mathfrak{G}%
\left( z\right) ^{2}}\tilde{F}_{\rho \sigma }\left( z\right) \right) \tilde{F%
}_{\alpha \beta }\left( z\right)  \notag \\
&+&\left. \frac{\partial \lambda \left( z\right) }{\partial \mathfrak{F}%
\left( z\right) }\left( \eta _{\alpha \rho }\eta _{\beta \sigma }-\eta
_{\alpha \sigma }\eta _{\beta \rho }\right) +\frac{\partial \lambda \left(
z\right) }{\partial \mathfrak{G}\left( z\right) }\varepsilon _{\alpha \beta
\rho \sigma }\right\} \delta \left( z-y\right) \delta \left( z-y^{\prime
}\right) \,;  \label{nm6.2}
\end{eqnarray}

\item Third functional derivatives:%
\begin{eqnarray}
&&\frac{\delta ^{3}\lambda \left( z\right) }{\delta F^{\alpha \beta }\left(
y\right) \delta F^{\rho \sigma }\left( y^{\prime }\right) \delta
F^{\varkappa \gamma }\left( y^{\prime \prime }\right) }= \\
&=&\frac{1}{8}\left\{ \left[ \frac{\partial ^{3}\lambda \left( z\right) }{%
\partial \mathfrak{F}^{2}\left( z\right) \partial \mathfrak{G}\left(
z\right) }F_{\varkappa \gamma }\left( z\right) +\frac{\partial ^{3}\lambda
\left( z\right) }{\partial \mathfrak{F}\left( z\right) \partial \mathfrak{G}%
^{2}\left( z\right) }\tilde{F}_{\varkappa \gamma }\left( z\right) \right] 
\tilde{F}_{\rho \sigma }\left( z\right) F_{\alpha \beta }\left( z\right)
\right.  \notag \\
&+&\frac{\partial ^{2}\lambda \left( z\right) }{\partial \mathfrak{F}\left(
z\right) \partial \mathfrak{G}\left( z\right) }\left[ \varepsilon
_{\varkappa \gamma \rho \sigma }F_{\alpha \beta }\left( z\right) +\left(
\eta _{\alpha \varkappa }\eta _{\beta \gamma }-\eta _{\alpha \gamma }\eta
_{\beta \varkappa }\right) \tilde{F}_{\rho \sigma }\left( z\right) \right] 
\notag \\
&+&\left[ \frac{\partial ^{3}\lambda \left( z\right) }{\partial \mathfrak{F}%
^{3}\left( z\right) }F_{\varkappa \gamma }\left( z\right) +\frac{\partial
^{3}\lambda \left( z\right) }{\partial \mathfrak{F}^{2}\left( z\right)
\partial \mathfrak{G}\left( z\right) }\tilde{F}_{\varkappa \gamma }\left(
z\right) \right] F_{\rho \sigma }\left( z\right) F_{\alpha \beta }\left(
z\right)  \notag \\
&+&\frac{\partial ^{2}\lambda \left( z\right) }{\partial \mathfrak{F}%
^{2}\left( z\right) }\left[ \left( \eta _{\gamma \sigma }\eta _{\rho
\varkappa }-\eta _{\gamma \rho }\eta _{\varkappa \sigma }\right) F_{\alpha
\beta }\left( z\right) +\left( \eta _{\alpha \varkappa }\eta _{\beta \gamma
}-\eta _{\alpha \gamma }\eta _{\beta \varkappa }\right) F_{\rho \sigma
}\left( z\right) \right]  \notag \\
&+&\left[ \frac{\partial ^{3}\lambda \left( z\right) }{\partial \mathfrak{G}%
\left( z\right) \partial \mathfrak{F}^{2}\left( z\right) }F_{\varkappa
\gamma }\left( z\right) +\frac{\partial ^{3}\lambda \left( z\right) }{%
\partial \mathfrak{G}^{2}\left( z\right) \partial \mathfrak{F}\left(
z\right) }\tilde{F}_{\varkappa \gamma }\left( z\right) \right] F_{\rho
\sigma }\left( z\right) \tilde{F}_{\alpha \beta }\left( z\right)  \notag \\
&+&\frac{\partial ^{2}\lambda \left( z\right) }{\partial \mathfrak{G}\left(
z\right) \partial \mathfrak{F}\left( z\right) }\left[ \left( \eta _{\gamma
\sigma }\eta _{\rho \varkappa }-\eta _{\gamma \rho }\eta _{\varkappa \sigma
}\right) \tilde{F}_{\alpha \beta }\left( z\right) +\varepsilon _{\varkappa
\gamma \alpha \beta }F_{\rho \sigma }\left( z\right) \right]  \notag \\
&+&\left[ \frac{\partial ^{3}\lambda \left( z\right) }{\partial \mathfrak{G}%
^{2}\left( z\right) \partial \mathfrak{F}\left( z\right) }F_{\varkappa
\gamma }\left( z\right) +\frac{\partial ^{3}\lambda \left( z\right) }{%
\partial \mathfrak{G}^{3}\left( z\right) }\tilde{F}_{\varkappa \gamma
}\left( z\right) \right] \tilde{F}_{\rho \sigma }\left( z\right) \tilde{F}%
_{\alpha \beta }\left( z\right)  \notag \\
&+&\frac{\partial ^{2}\lambda \left( z\right) }{\partial \mathfrak{G}%
^{2}\left( z\right) }\left[ \varepsilon _{\varkappa \gamma \rho \sigma }%
\tilde{F}_{\alpha \beta }\left( z\right) +\varepsilon _{\varkappa \gamma
\alpha \beta }\tilde{F}_{\rho \sigma }\left( z\right) \right]  \notag \\
&+&\left[ \frac{\partial ^{2}\lambda \left( z\right) }{\partial \mathfrak{F}%
^{2}\left( z\right) }F_{\varkappa \gamma }\left( z\right) +\frac{\partial
^{2}\lambda \left( z\right) }{\partial \mathfrak{F}\left( z\right) \partial 
\mathfrak{G}\left( z\right) }\tilde{F}_{\varkappa \gamma }\left( z\right) %
\right] \left( \eta _{\alpha \rho }\eta _{\beta \sigma }-\eta _{\alpha
\sigma }\eta _{\beta \rho }\right)  \notag \\
&+&\left. \left[ \frac{\partial ^{2}\lambda \left( z\right) }{\partial 
\mathfrak{G}^{2}\left( z\right) }\tilde{F}_{\varkappa \gamma }\left(
z\right) +\frac{\partial ^{2}\lambda \left( z\right) }{\partial \mathfrak{G}%
\left( z\right) \partial \mathfrak{F}\left( z\right) }F_{\varkappa \gamma
}\left( z\right) \right] \varepsilon _{\alpha \beta \rho \sigma }\right\}
\delta \left( z-y\right) \delta \left( z-y^{\prime }\right) \delta \left(
z-y^{\prime \prime }\right) \,.  \label{nm6.4}
\end{eqnarray}
\end{itemize}

With the help of the results above one may finally write nonlinear Maxwell
equations by truncating the series (\ref{nm5.1}), (\ref{nm5.2}) to the third
power in the deviations $f^{\alpha \beta }\left( x\right) $. To this aim we
simplify the notation by labeling the partial derivatives of the effective
Lagrangian as follows:%
\begin{eqnarray}
&&\lambda =\mathfrak{L}_{\mathfrak{F}}\,,\ \ \tilde{\lambda}=\mathfrak{L}_{%
\mathfrak{G}}\,,\ \ \frac{\partial \lambda }{\partial \mathfrak{F}}=%
\mathfrak{L}_{\mathfrak{FF}}\,,  \notag \\
&&\frac{\partial \tilde{\lambda}}{\partial \mathfrak{G}}=\mathfrak{L}_{%
\mathfrak{GG}}\,,\ \ \frac{\partial \lambda }{\partial \mathfrak{G}}=%
\mathfrak{L}_{\mathfrak{FG}}=\frac{\partial \tilde{\lambda}}{\partial 
\mathfrak{F}}\,,  \notag \\
&&\frac{\partial ^{2}\lambda }{\partial \mathfrak{G}^{2}}=\mathfrak{L}_{%
\mathfrak{FGG}}=\frac{\partial ^{2}\tilde{\lambda}}{\partial \mathfrak{F}%
\partial \mathfrak{G}}\,,\ \ \frac{\partial ^{2}\lambda }{\partial \mathfrak{%
F}\partial \mathfrak{G}}=\mathfrak{L}_{\mathfrak{FFG}}=\frac{\partial ^{2}%
\tilde{\lambda}}{\partial \mathfrak{F}^{2}}\,,  \notag \\
&&\frac{\partial ^{2}\lambda }{\partial \mathfrak{F}^{2}}=\mathfrak{L}_{%
\mathfrak{FFF}}\,,\ \ \frac{\partial ^{2}\tilde{\lambda}}{\partial \mathfrak{%
G}^{2}}=\mathfrak{L}_{\mathfrak{GGG}}\,,\ \ \frac{\partial ^{3}\lambda }{%
\partial \mathfrak{F}^{3}}=\mathfrak{L}_{\mathfrak{FFFF}}\,,  \notag \\
&&\frac{\partial ^{3}\lambda }{\partial \mathfrak{G}^{3}}=\mathfrak{L}_{%
\mathfrak{FGGG}}=\frac{\partial ^{3}\tilde{\lambda}}{\partial \mathfrak{F}%
^{3}}\,,\ \ \frac{\partial ^{3}\lambda }{\partial \mathfrak{F}^{2}\partial 
\mathfrak{G}}=\mathfrak{L}_{\mathfrak{FFFG}}=\frac{\partial ^{3}\tilde{%
\lambda}}{\partial \mathfrak{F}\partial \mathfrak{G}^{2}}\,,  \notag \\
&&\frac{\partial ^{3}\lambda }{\partial \mathfrak{F}\partial \mathfrak{G}^{2}%
}=\mathfrak{L}_{\mathfrak{FFGG}}=\frac{\partial ^{3}\tilde{\lambda}}{%
\partial \mathfrak{F}^{2}\partial \mathfrak{G}}\,,\ \ \frac{\partial ^{3}%
\tilde{\lambda}\left( z\right) }{\partial \mathfrak{G}^{3}\left( z\right) }=%
\mathfrak{L}_{\mathfrak{GGGG}}\,.  \label{nm6.3.1}
\end{eqnarray}%
Hence the first functional derivatives of $\lambda \left( z\right) $ and $%
\tilde{\lambda}\left( z\right) $ (\ref{nm6.1}) (as they should be used in (%
\ref{nm5.1}), (\ref{nm5.2})) are given by,%
\begin{eqnarray}
&&\left. \frac{\delta \lambda \left( z\right) }{\delta F^{\alpha \beta
}\left( y\right) }\right\vert _{f=0}=\frac{1}{2}\left( \mathfrak{L}_{%
\mathfrak{FF}}\mathcal{F}_{\alpha \beta }+\mathfrak{L}_{\mathfrak{FG}}%
\mathcal{\tilde{F}}_{\alpha \beta }\right) \delta \left( z-y\right) \,, 
\notag \\
&&\left. \frac{\delta \tilde{\lambda}\left( z\right) }{\delta F^{\alpha
\beta }\left( y\right) }\right\vert _{f=0}=\frac{1}{2}\left( \mathfrak{L}_{%
\mathfrak{FG}}\mathcal{F}_{\alpha \beta }+\mathfrak{L}_{\mathfrak{GG}}%
\mathcal{\tilde{F}}_{\alpha \beta }\right) \delta \left( z-y\right) \,.
\label{nm7.1}
\end{eqnarray}%
Next, the second derivatives (\ref{nm6.2}) are%
\begin{eqnarray}
\left. \frac{\delta ^{2}\lambda \left( z\right) }{\delta F^{\alpha \beta
}\left( y\right) \delta F^{\rho \sigma }\left( y^{\prime }\right) }%
\right\vert _{f=0} &=&\frac{1}{4}\left\{ \left[ \mathfrak{L}_{\mathfrak{FFG}}%
\mathcal{\tilde{F}}_{\rho \sigma }+\mathfrak{L}_{\mathfrak{FFF}}\mathcal{F}%
_{\rho \sigma }\right] \mathcal{F}_{\alpha \beta }+\left[ \mathfrak{L}_{%
\mathfrak{FFG}}\mathcal{F}_{\rho \sigma }+\mathfrak{L}_{\mathfrak{FGG}}%
\mathcal{\tilde{F}}_{\rho \sigma }\right] \mathcal{\tilde{F}}_{\alpha \beta
}\right.  \notag \\
&+&\left. \mathfrak{L}_{\mathfrak{FF}}\left( \eta _{\alpha \rho }\eta
_{\beta \sigma }-\eta _{\alpha \sigma }\eta _{\beta \rho }\right) +\mathfrak{%
L}_{\mathfrak{FG}}\varepsilon _{\alpha \beta \rho \sigma }\right\} \delta
\left( z-y\right) \delta \left( z-y^{\prime }\right) \,,  \notag \\
\left. \frac{\delta ^{2}\tilde{\lambda}\left( z\right) }{\delta F^{\alpha
\beta }\left( y\right) \delta F^{\rho \sigma }\left( y^{\prime }\right) }%
\right\vert _{f=0} &=&\frac{1}{4}\left\{ \left[ \mathfrak{L}_{\mathfrak{FGG}}%
\mathcal{\tilde{F}}_{\rho \sigma }+\mathfrak{L}_{\mathfrak{FFG}}\mathcal{F}%
_{\rho \sigma }\right] \mathcal{F}_{\alpha \beta }+\left[ \mathfrak{L}_{%
\mathfrak{FGG}}\mathcal{F}_{\rho \sigma }+\mathfrak{L}_{\mathfrak{GGG}}%
\mathcal{\tilde{F}}_{\rho \sigma }\right] \mathcal{\tilde{F}}_{\alpha \beta
}\right.  \notag \\
&+&\left. \mathfrak{L}_{\mathfrak{FG}}\left( \eta _{\alpha \rho }\eta
_{\beta \sigma }-\eta _{\alpha \sigma }\eta _{\beta \rho }\right) +\mathfrak{%
L}_{\mathfrak{GG}}\varepsilon _{\alpha \beta \rho \sigma }\right\} \delta
\left( z-y\right) \delta \left( z-y^{\prime }\right) \,,  \label{nm7.2}
\end{eqnarray}%
and, at last, the third-order derivatives (\ref{nm6.4}) are reduced to%
\begin{eqnarray}
\left. \frac{\delta ^{3}\lambda \left( z\right) }{\delta F^{\alpha \beta
}\left( y\right) \delta F^{\rho \sigma }\left( y^{\prime }\right) \delta
F^{\varkappa \gamma }\left( y^{\prime \prime }\right) }\right\vert _{f=0} &=&%
\frac{1}{8}\left\{ \left[ \mathfrak{L}_{\mathfrak{FFFG}}\mathcal{F}%
_{\varkappa \gamma }+\mathfrak{L}_{\mathfrak{FFGG}}\mathcal{\tilde{F}}%
_{\varkappa \gamma }\right] \mathcal{\tilde{F}}_{\rho \sigma }\mathcal{F}%
_{\alpha \beta }\right.  \notag \\
&+&\mathfrak{L}_{\mathfrak{FFG}}\left[ \varepsilon _{\varkappa \gamma \rho
\sigma }\mathcal{F}_{\alpha \beta }+\left( \eta _{\alpha \varkappa }\eta
_{\beta \gamma }-\eta _{\alpha \gamma }\eta _{\beta \varkappa }\right) 
\mathcal{\tilde{F}}_{\rho \sigma }\right]  \notag \\
&+&\left[ \mathfrak{L}_{\mathfrak{FFFF}}\mathcal{F}_{\varkappa \gamma }+%
\mathfrak{L}_{\mathfrak{FFFG}}\mathcal{\tilde{F}}_{\varkappa \gamma }\right] 
\mathcal{F}_{\rho \sigma }\mathcal{F}_{\alpha \beta }  \notag \\
&+&\mathfrak{L}_{\mathfrak{FFF}}\left[ \left( \eta _{\gamma \sigma }\eta
_{\rho \varkappa }-\eta _{\gamma \rho }\eta _{\varkappa \sigma }\right) 
\mathcal{F}_{\alpha \beta }+\left( \eta _{\alpha \varkappa }\eta _{\beta
\gamma }-\eta _{\alpha \gamma }\eta _{\beta \varkappa }\right) \mathcal{F}%
_{\rho \sigma }\right]  \notag \\
&+&\left[ \mathfrak{L}_{\mathfrak{FFFG}}\mathcal{F}_{\varkappa \gamma }+%
\mathfrak{L}_{\mathfrak{FFGG}}\mathcal{\tilde{F}}_{\varkappa \gamma }\right] 
\mathcal{F}_{\rho \sigma }\mathcal{\tilde{F}}_{\alpha \beta }  \notag \\
&+&\mathfrak{L}_{\mathfrak{FFG}}\left[ \left( \eta _{\gamma \sigma }\eta
_{\rho \varkappa }-\eta _{\gamma \rho }\eta _{\varkappa \sigma }\right) 
\mathcal{\tilde{F}}_{\alpha \beta }+\varepsilon _{\varkappa \gamma \alpha
\beta }\mathcal{F}_{\rho \sigma }\right]  \notag \\
&+&\left[ \mathfrak{L}_{\mathfrak{FFGG}}\mathcal{F}_{\varkappa \gamma }+%
\mathfrak{L}_{\mathfrak{FGGG}}\mathcal{\tilde{F}}_{\varkappa \gamma }\right] 
\mathcal{\tilde{F}}_{\rho \sigma }\mathcal{\tilde{F}}_{\alpha \beta }+%
\mathfrak{L}_{\mathfrak{FGG}}\left[ \varepsilon _{\varkappa \gamma \rho
\sigma }\mathcal{\tilde{F}}_{\alpha \beta }+\varepsilon _{\varkappa \gamma
\alpha \beta }\mathcal{\tilde{F}}_{\rho \sigma }\right]  \notag \\
&+&\left[ \mathfrak{L}_{\mathfrak{FFF}}\mathcal{F}_{\varkappa \gamma }+%
\mathfrak{L}_{\mathfrak{FFG}}\mathcal{\tilde{F}}_{\varkappa \gamma }\right]
\left( \eta _{\alpha \rho }\eta _{\beta \sigma }-\eta _{\alpha \sigma }\eta
_{\beta \rho }\right)  \notag \\
&+&\left. \left[ \mathfrak{L}_{\mathfrak{FGG}}\mathcal{\tilde{F}}_{\varkappa
\gamma }+\mathfrak{L}_{\mathfrak{FFG}}\mathcal{F}_{\varkappa \gamma }\right]
\varepsilon _{\alpha \beta \rho \sigma }\right\} \delta \left( z-y\right)
\delta \left( z-y^{\prime }\right) \delta \left( z-y^{\prime \prime }\right)
\,,  \notag \\
\left. \frac{\delta ^{3}\tilde{\lambda}\left( z\right) }{\delta F^{\alpha
\beta }\left( y\right) \delta F^{\rho \sigma }\left( y^{\prime }\right)
\delta F^{\varkappa \gamma }\left( y^{\prime \prime }\right) }\right\vert
_{f=0} &=&\frac{1}{8}\left\{ \left[ \mathfrak{L}_{\mathfrak{FFGG}}\mathcal{F}%
_{\varkappa \gamma }+\mathfrak{L}_{\mathfrak{FGGG}}\mathcal{\tilde{F}}%
_{\varkappa \gamma }\right] \mathcal{\tilde{F}}_{\rho \sigma }\mathcal{F}%
_{\alpha \beta }\right.  \notag \\
&+&\mathfrak{L}_{\mathfrak{FGG}}\left[ \varepsilon _{\varkappa \gamma \rho
\sigma }\mathcal{F}_{\alpha \beta }+\left( \eta _{\alpha \varkappa }\eta
_{\beta \gamma }-\eta _{\alpha \gamma }\eta _{\beta \varkappa }\right) 
\mathcal{\tilde{F}}_{\rho \sigma }\right]  \notag \\
&+&\left[ \mathfrak{L}_{\mathfrak{FFFG}}\mathcal{F}_{\varkappa \gamma }+%
\mathfrak{L}_{\mathfrak{FFGG}}\mathcal{\tilde{F}}_{\varkappa \gamma }\right] 
\mathcal{F}_{\rho \sigma }\mathcal{F}_{\alpha \beta }  \notag \\
&+&\mathfrak{L}_{\mathfrak{FFG}}\left[ \left( \eta _{\gamma \sigma }\eta
_{\rho \varkappa }-\eta _{\gamma \rho }\eta _{\varkappa \sigma }\right) 
\mathcal{F}_{\alpha \beta }+\left( \eta _{\alpha \varkappa }\eta _{\beta
\gamma }-\eta _{\alpha \gamma }\eta _{\beta \varkappa }\right) \mathcal{F}%
_{\rho \sigma }\right]  \notag \\
&+&\left[ \mathfrak{L}_{\mathfrak{FFGG}}\mathcal{F}_{\varkappa \gamma }+%
\mathfrak{L}_{\mathfrak{FGGG}}\mathcal{\tilde{F}}_{\varkappa \gamma }\right] 
\mathcal{F}_{\rho \sigma }\mathcal{\tilde{F}}_{\alpha \beta }  \notag \\
&+&\mathfrak{L}_{\mathfrak{FGG}}\left[ \left( \eta _{\gamma \sigma }\eta
_{\rho \varkappa }-\eta _{\gamma \rho }\eta _{\varkappa \sigma }\right) 
\mathcal{\tilde{F}}_{\alpha \beta }+\varepsilon _{\varkappa \gamma \alpha
\beta }\mathcal{F}_{\rho \sigma }\right]  \notag \\
&+&\left[ \mathfrak{L}_{\mathfrak{FGGG}}\mathcal{F}_{\varkappa \gamma }+%
\mathfrak{L}_{\mathfrak{GGGG}}\mathcal{\tilde{F}}_{\varkappa \gamma }\right] 
\mathcal{\tilde{F}}_{\rho \sigma }\mathcal{\tilde{F}}_{\alpha \beta }+%
\mathfrak{L}_{\mathfrak{GGG}}\left[ \varepsilon _{\varkappa \gamma \rho
\sigma }\mathcal{\tilde{F}}_{\alpha \beta }+\varepsilon _{\varkappa \gamma
\alpha \beta }\mathcal{\tilde{F}}_{\rho \sigma }\right]  \notag \\
&+&\left[ \mathfrak{L}_{\mathfrak{FFG}}\mathcal{F}_{\varkappa \gamma }+%
\mathfrak{L}_{\mathfrak{FGG}}\mathcal{\tilde{F}}_{\varkappa \gamma }\right]
\left( \eta _{\alpha \rho }\eta _{\beta \sigma }-\eta _{\alpha \sigma }\eta
_{\beta \rho }\right)  \notag \\
&+&\left. \left[ \mathfrak{L}_{\mathfrak{GGG}}\mathcal{\tilde{F}}_{\varkappa
\gamma }+\mathfrak{L}_{\mathfrak{FGG}}\mathcal{F}_{\varkappa \gamma }\right]
\varepsilon _{\alpha \beta \rho \sigma }\right\} \delta \left( z-y\right)
\delta \left( z-y^{\prime }\right) \delta \left( z-y^{\prime \prime }\right)
\,.  \label{nm7.2.2}
\end{eqnarray}%
Here it is meant that the derivatives (\ref{nm6.3.1}) are reduced to the
background field $\mathcal{F}^{\alpha \beta }\left( x\right) $.

Truncating the series in the first power of deviations $f^{\alpha \beta
}\left( x\right) $, one obtains the nonlinear Maxwell equations
corresponding to the linear response of the background field applied to a
small source $j^{\mu }\left( x\right) $. As discussed at the Subsec. \ref%
{Expansion}, the nonlinear current (\ref{nm8.1}) takes the form%
\begin{eqnarray}
j_{\mu }^{\text{\textrm{nl}}}\left( x\right) &=&\partial ^{\tau }\left(
\lambda _{\tau \mu }\left( x\right) +\tilde{\lambda}_{\tau \mu }\left(
x\right) \right) =j_{\mu }^{\text{\textrm{lin}}}\left( x\right) \,,  \notag
\\
j_{\mu }^{\text{\textrm{lin}}}\left( x\right) &=&\mathfrak{L}_{\mathfrak{F}%
}\partial ^{\tau }f_{\tau \mu }\left( x\right) +\frac{1}{2}\left( \mathfrak{L%
}_{\mathfrak{FF}}\mathcal{F}_{\alpha \beta }+\mathfrak{L}_{\mathfrak{FG}}%
\widetilde{\mathcal{F}}_{\alpha \beta }\right) \mathcal{F}_{\tau \mu
}\partial ^{\tau }f^{\alpha \beta }\left( x\right)  \notag \\
&+&\mathfrak{L}_{\mathfrak{G}}\partial ^{\tau }\tilde{f}_{\tau \mu }\left(
x\right) +\frac{1}{2}\left( \mathfrak{L}_{\mathfrak{FG}}\mathcal{F}_{\alpha
\beta }+\mathfrak{L}_{\mathfrak{GG}}\widetilde{\mathcal{F}}_{\alpha \beta
}\right) \widetilde{\mathcal{F}}_{\tau \mu }\partial ^{\tau }f^{\alpha \beta
}\left( x\right) \,.  \label{nm9.1}
\end{eqnarray}%
Truncating the series in the second power of deviations, the nonlinear
current (\ref{nonlincurrent}) now has more terms, and can be splited in two
parts,%
\begin{equation}
j_{\mu }^{\text{\textrm{nl}}}\left( x\right) \simeq j_{\mu }^{\text{\textrm{%
lin}}}\left( x\right) +j_{\mu }^{\text{\textrm{qua}}}\left( x\right) \,,
\label{nm9.2}
\end{equation}%
where $j_{\mu }^{\text{\textrm{lin}}}\left( x\right) $ is linear in $%
f^{\alpha \beta }\left( x\right) $ while $j_{\mu }^{\text{\textrm{qua}}%
}\left( x\right) $ is quadratic. Again when the background is formed by
constant fields, $j_{\mu }^{\text{\textrm{lin}}}\left( x\right) $ takes the
same form as (\ref{nm9.1}) while the quadratic part read%
\begin{eqnarray}
j_{\mu }^{\text{\textrm{qua}}}\left( x\right) &=&\frac{1}{8}\left( \mathfrak{%
L}_{\mathfrak{FFG}}\mathcal{\tilde{F}}_{\rho \sigma }+\mathfrak{L}_{%
\mathfrak{FFF}}\mathcal{F}_{\rho \sigma }\right) \mathcal{F}_{\tau \mu }%
\mathcal{F}_{\alpha \beta }\partial ^{\tau }\left[ f^{\alpha \beta }\left(
x\right) f^{\rho \sigma }\left( x\right) \right]  \notag \\
&+&\frac{1}{8}\left( \mathfrak{L}_{\mathfrak{FFG}}\mathcal{F}_{\rho \sigma }+%
\mathfrak{L}_{\mathfrak{FGG}}\mathcal{\tilde{F}}_{\rho \sigma }\right) 
\mathcal{F}_{\tau \mu }\mathcal{\tilde{F}}_{\alpha \beta }\partial ^{\tau }%
\left[ f^{\alpha \beta }\left( x\right) f^{\rho \sigma }\left( x\right) %
\right]  \notag \\
&+&\frac{1}{4}\mathfrak{L}_{\mathfrak{FF}}\mathcal{F}_{\tau \mu }\partial
^{\tau }\left[ f_{\rho \sigma }\left( x\right) f^{\rho \sigma }\left(
x\right) \right] +\frac{1}{4}\mathfrak{L}_{\mathfrak{FG}}\mathcal{F}_{\tau
\mu }\partial ^{\tau }\left[ \tilde{f}_{\rho \sigma }\left( x\right) f^{\rho
\sigma }\left( x\right) \right]  \notag \\
&+&\frac{1}{2}\left( \mathfrak{L}_{\mathfrak{FF}}\mathcal{F}_{\alpha \beta }+%
\mathfrak{L}_{\mathfrak{FG}}\mathcal{\tilde{F}}_{\alpha \beta }\right)
\partial ^{\tau }\left[ f^{\alpha \beta }\left( x\right) f_{\tau \mu }\left(
x\right) \right]  \notag \\
&+&\frac{1}{8}\left( \mathfrak{L}_{\mathfrak{FGG}}\mathcal{\tilde{F}}_{\rho
\sigma }+\mathfrak{L}_{\mathfrak{FFG}}\mathcal{F}_{\rho \sigma }\right) 
\mathcal{F}_{\alpha \beta }\mathcal{\tilde{F}}_{\tau \mu }\partial ^{\tau }%
\left[ f^{\alpha \beta }\left( x\right) f^{\rho \sigma }\left( x\right) %
\right]  \notag \\
&+&\frac{1}{8}\left( \mathfrak{L}_{\mathfrak{FGG}}\mathcal{F}_{\rho \sigma }+%
\mathfrak{L}_{\mathfrak{GGG}}\mathcal{\tilde{F}}_{\rho \sigma }\right) 
\mathcal{\tilde{F}}_{\alpha \beta }\mathcal{\tilde{F}}_{\tau \mu }\partial
^{\tau }\left[ f^{\alpha \beta }\left( x\right) f^{\rho \sigma }\left(
x\right) \right]  \notag \\
&+&\frac{1}{4}\mathfrak{L}_{\mathfrak{FG}}\mathcal{\tilde{F}}_{\tau \mu
}\partial ^{\tau }\left[ f_{\rho \sigma }\left( x\right) f^{\rho \sigma
}\left( x\right) \right] +\frac{1}{4}\mathfrak{L}_{\mathfrak{GG}}\mathcal{%
\tilde{F}}_{\tau \mu }\partial ^{\tau }\left[ \tilde{f}_{\rho \sigma }\left(
x\right) f^{\rho \sigma }\left( x\right) \right]  \notag \\
&+&\frac{1}{2}\left( \mathfrak{L}_{\mathfrak{FG}}\mathcal{F}_{\alpha \beta }+%
\mathfrak{L}_{\mathfrak{GG}}\mathcal{\tilde{F}}_{\alpha \beta }\right)
\partial ^{\tau }\left[ \tilde{f}_{\tau \mu }\left( x\right) f^{\alpha \beta
}\left( x\right) \right] \,.  \label{nm9.3}
\end{eqnarray}

Following the procedure discussed above, one may use the formulas (\ref%
{nm5.5}), (\ref{nm5.6}), (\ref{nm6.4}), (\ref{nm7.2.2}) and construct the
cubic nonlinear current. Although its exact expression can be rather
complicated, it takes a simple form in the vacuum. To derive such an
equation one must set $\mathcal{F}^{\alpha \beta }=0$ in all formulas above
such that only cubic terms survives. For example in this case the tensor
coefficients (\ref{nm4new}) reads%
\begin{eqnarray}
&&\left. \lambda _{\tau \mu }\left( z\right) \right\vert _{\mathcal{F}%
=0}\simeq \left[ \mathcal{L}_{\mathfrak{FF}}\mathfrak{f}\left( z\right) +%
\mathcal{L}_{\mathfrak{FG}}\mathfrak{g}\left( z\right) \right] f_{\tau \mu
}\left( z\right) \,,  \notag \\
&&\left. \tilde{\lambda}_{\tau \mu }\left( z\right) \right\vert _{\mathcal{F}%
=0}\simeq \left[ \mathcal{L}_{\mathfrak{FG}}\mathfrak{f}\left( z\right) +%
\mathcal{L}_{\mathfrak{GG}}\mathfrak{g}\left( z\right) \right] \tilde{f}%
_{\tau \mu }\left( z\right) \,,  \label{nm8.5} \\
&&\mathfrak{f}\left( z\right) =\frac{1}{4}f_{\alpha \beta }\left( z\right)
f^{\alpha \beta }\left( z\right) \,,\ \ \mathfrak{g}\left( z\right) =\frac{1%
}{4}f_{\alpha \beta }\left( z\right) \tilde{f}^{\alpha \beta }\left(
z\right) \,,  \notag
\end{eqnarray}%
Then using the results above the nonlinear Maxwell equations in the vacuum
takes the form%
\begin{equation}
\partial ^{\nu }f_{\nu \mu }=\mathcal{L}_{\mathfrak{FF}}\partial ^{\tau
}\left( \mathfrak{f}f_{\tau \mu }\right) +\mathcal{L}_{\mathfrak{FG}%
}\partial ^{\tau }\left( \mathfrak{f}\tilde{f}_{\tau \mu }+\mathfrak{g}%
f_{\tau \mu }\right) +\mathcal{L}_{\mathfrak{GG}}\partial ^{\tau }\left( 
\mathfrak{g}\tilde{f}_{\tau \mu }\right) \,.  \label{nm9.4}
\end{equation}%
This equation has been previously derived following a different method. See 
\cite{CaiGitSha2013}.

Restricting to the static case where the time derivatives are absent, the
zero component of (\ref{nm9.1}) $\left( \mu =0\right) $, gives the nonlinear
nonhomogeneous Maxwell equation for the electric field%
\begin{equation}
\boldsymbol{\nabla }\cdot \mathbf{E}=\boldsymbol{\nabla }\cdot \left[ 
\mathcal{L}_{\mathfrak{FF}}\mathfrak{f}\mathbf{E}+\mathcal{L}_{\mathfrak{FG}%
}\left( \mathfrak{f}\mathbf{B}+\mathfrak{g}\mathbf{E}\right) +\mathcal{L}_{%
\mathfrak{GG}}\mathfrak{g}\mathbf{B}\right] \,,  \label{nm9a}
\end{equation}%
while the spacial component $\left( \mu =1,2,3\right) $, provides the
nonlinear nonhomogeneous Maxwell equation for the magnetic field%
\begin{equation}
\boldsymbol{\nabla }\times \mathbf{B}=\boldsymbol{\nabla }\times \left[ 
\mathcal{L}_{\mathfrak{FF}}\mathfrak{f}\mathbf{B}-\mathcal{L}_{\mathfrak{FG}%
}\left( \mathfrak{f}\mathbf{E}-\mathfrak{g}\mathbf{B}\right) -\mathcal{L}_{%
\mathfrak{GG}}\mathfrak{g}\mathbf{E}\right] \,.  \label{nm9b}
\end{equation}

\section{Projection operator\label{AppII}}

In this Appendix we evaluate the action of the projection operator (\ref%
{sc2.5c}). Substituting (\ref{in2}) in (\ref{nnew13a}) the auxiliary
electric field $\mathbf{\mathcal{E}}\left( \mathbf{x}\right) $ (\ref{nnew13a}%
) takes the form%
\begin{eqnarray}
\mathbf{\mathcal{E}}\left( \mathbf{x}\right) &=&\frac{q}{4\pi }\mathfrak{L}_{%
\mathfrak{F}}\Phi \left( r\right) \mathbf{x}-\frac{q}{4\pi }\Phi \left(
r\right) \left( \mathfrak{L}_{\mathfrak{FF}}\mathbf{\overline{E}}+\mathfrak{L%
}_{\mathfrak{FG}}\mathbf{\overline{B}}\right) \left( \mathbf{\overline{E}}%
\cdot \mathbf{x}\right)  \notag \\
&-&\frac{q}{4\pi }\Phi \left( r\right) \left( \mathfrak{L}_{\mathfrak{FG}}%
\mathbf{\overline{E}}+\mathfrak{L}_{\mathfrak{GG}}\mathbf{\overline{B}}%
\right) \left( \mathbf{\overline{B}}\cdot \mathbf{x}\right) \,.  \label{arb1}
\end{eqnarray}%
In accordance to (\ref{sc2.5c}), the integral of the auxiliary electric
field above can be written as,%
\begin{eqnarray}
&&\int d\mathbf{y}\frac{\mathcal{E}^{j}\left( \mathbf{y}\right) }{\left\vert 
\mathbf{x-y}\right\vert }=\frac{q}{4\pi }\mathfrak{L}_{\mathfrak{F}}\mathcal{%
I}^{j}\left( \mathbf{x}\right) +\mathfrak{T}_{E}^{jk}\mathcal{I}^{k}\left( 
\mathbf{x}\right) \,,  \notag \\
&&\mathcal{I}^{j}\left( \mathbf{x}\right) =\int d\mathbf{y}\frac{\Phi \left(
y\right) y^{j}}{\left\vert \mathbf{x-y}\right\vert }\,,  \notag \\
&&\mathfrak{T}_{E}^{jk}=-\frac{q}{4\pi }\left[ \left( \mathfrak{L}_{%
\mathfrak{FF}}\overline{E}^{j}+\mathfrak{L}_{\mathfrak{FG}}\overline{B}%
^{j}\right) \overline{E}^{k}+\left( \mathfrak{L}_{\mathfrak{FG}}\overline{E}%
^{j}+\mathfrak{L}_{\mathfrak{GG}}\overline{B}^{j}\right) \overline{B}^{k}%
\right] \,,  \label{arb5}
\end{eqnarray}%
where $\mathcal{I}^{j}\left( \mathbf{x}\right) $ can be expressed in terms
of the auxiliary function $\varrho \left( r\right) $,%
\begin{equation*}
\mathcal{I}^{j}\left( \mathbf{x}\right) =2\pi \varrho \left( r\right)
x^{j}\,,\ \ 2\pi \varrho \left( r\right) =\frac{1}{r^{2}}\int d\mathbf{y}%
\frac{\Phi \left( y\right) \left( \mathbf{y\cdot x}\right) }{\left\vert 
\mathbf{x-y}\right\vert }\,,
\end{equation*}%
whose explicit form reads, 
\begin{equation}
\varrho \left( r\right) =\frac{1}{R}\left( 1-\frac{r^{2}}{5R^{2}}\right)
\theta \left( R-r\right) +\frac{1}{r}\left( 1-\frac{R^{2}}{5r^{2}}\right)
\theta \left( r-R\right) \,.  \label{arb7}
\end{equation}

Using this result one may write the following set of relations:%
\begin{eqnarray}
&&\partial _{i}\partial _{j}\mathcal{I}^{j}\left( \mathbf{x}\right) =\frac{%
2\pi x^{i}}{r}\left[ 4\varrho ^{\prime }\left( r\right) +r\varrho ^{\prime
\prime }\left( r\right) \right] \,,  \notag \\
&&\partial _{i}\partial _{j}\mathcal{I}^{k}\left( \mathbf{x}\right) =2\pi %
\left[ \left( \delta ^{ij}x^{k}+x^{j}\delta ^{ik}+x^{i}\delta ^{jk}-\frac{%
x^{j}x^{k}x^{i}}{r^{2}}\right) \frac{\varrho ^{\prime }\left( r\right) }{r}+%
\frac{x^{j}x^{k}x^{i}}{r^{2}}\varrho ^{\prime \prime }\left( r\right) \right]
\,,  \label{arb8}
\end{eqnarray}%
where the primes denote differentiations with respect to $r$, and%
\begin{eqnarray}
&&\mathfrak{T}_{E}^{ik}x^{k}=-\frac{q}{4\pi }\left( V_{\mathfrak{F}}%
\overline{E}^{i}+V_{\mathfrak{G}}\overline{B}^{i}\right) =\mathfrak{T}%
_{E}^{ki}x^{k}\,,  \notag \\
&&V_{\mathfrak{F}}=\mathfrak{L}_{\mathfrak{FF}}\left( \mathbf{\overline{E}}%
\cdot \mathbf{x}\right) +\mathfrak{L}_{\mathfrak{FG}}\left( \mathbf{%
\overline{B}}\cdot \mathbf{x}\right) \,,\ \ V_{\mathfrak{G}}=\mathfrak{L}_{%
\mathfrak{FG}}\left( \mathbf{\overline{E}}\cdot \mathbf{x}\right) +\mathfrak{%
L}_{\mathfrak{GG}}\left( \mathbf{\overline{B}}\cdot \mathbf{x}\right) \,, 
\notag \\
&&\mathfrak{T}_{E}^{jk}x^{i}\delta ^{jk}=\mathfrak{T}_{E}^{jj}x^{i}=\frac{q}{%
4\pi }\tilde{b}x^{i}\,,\ \ \tilde{b}=-\left[ \mathfrak{L}_{\mathfrak{FF}}%
\mathbf{\overline{E}}^{2}+\mathfrak{L}_{\mathfrak{GG}}\mathbf{\overline{B}}%
^{2}-2\mathfrak{GL}_{\mathfrak{FG}}\right] \,,  \notag \\
&&x^{j}\mathfrak{T}_{E}^{jk}x^{k}=-\frac{q}{4\pi }\left[ \mathfrak{L}_{%
\mathfrak{FF}}\left( \mathbf{\overline{E}}\cdot \mathbf{x}\right) ^{2}+%
\mathfrak{L}_{\mathfrak{GG}}\left( \mathbf{\overline{B}}\cdot \mathbf{x}%
\right) ^{2}+2\mathfrak{L}_{\mathfrak{FG}}\left( \mathbf{\overline{B}}\cdot 
\mathbf{x}\right) \left( \mathbf{\overline{E}}\cdot \mathbf{x}\right) \right]
\,,  \label{arb9}
\end{eqnarray}

With the help of (\ref{arb8}), (\ref{arb9}), one sees finally that the
action of partial derivatives on (\ref{arb5}) has the result%
\begin{eqnarray}
\frac{\partial _{i}\partial _{j}}{4\pi }\int d\mathbf{y}\frac{\mathcal{E}%
^{j}\left( \mathbf{y}\right) }{\left\vert \mathbf{x-y}\right\vert } &=&\frac{%
q}{8\pi }\mathfrak{L}_{\mathfrak{F}}\left( \frac{4\varrho ^{\prime }\left(
r\right) }{r}+\varrho ^{\prime \prime }\left( r\right) \right) x^{i}  \notag
\\
&+&\left( 2\mathfrak{T}_{E}^{jk}\delta ^{ij}x^{k}+\mathfrak{T}%
_{E}^{jj}x^{i}\right) \frac{\varrho ^{\prime }\left( r\right) }{2r}+\frac{%
x^{j}\mathfrak{T}_{E}^{jk}x^{k}x^{i}}{2r^{2}}\left( \varrho ^{\prime \prime
}\left( r\right) -\frac{\varrho ^{\prime }\left( r\right) }{r}\right) \,.
\label{arb12}
\end{eqnarray}%
Since the function $\varrho \left( r\right) $ as given by (\ref{arb7}), and
its derivative $\varrho ^{\prime }\left( r\right) $ are both continuous in
the point $r=R$\ one must not differentiate the step functions $\theta
\left( R-r\right) $ and $\theta \left( r-R\right) $\ when calculating (\ref%
{arb12}). Then the latter produces Eq. (\ref{arb14}).

\end{appendices}

\end{document}